\documentclass{sig-alternate-05-2015}
\usepackage{times}
\usepackage{balance}
\usepackage{url}
\usepackage{epsfig}
\usepackage{graphicx}
\usepackage{subfigure}
\usepackage{amsmath}
\usepackage{amssymb} 
\usepackage{amsfonts}
\usepackage{color}
\usepackage{multirow}
\usepackage{algorithm}
\usepackage[utf8]{inputenc}
\usepackage[T1]{fontenc}
\usepackage{cases}
\usepackage{capt-of}

\newenvironment{packed_itemize}{
\begin{itemize}
\setlength{\itemsep}{1pt}
\setlength{\parskip}{0pt}
\setlength{\parsep}{0pt}
\setlength{\headsep}{0pt}
\setlength{\topskip}{0pt}
\setlength{\topmargin}{0pt}
\setlength{\topsep}{0pt}
\setlength{\partopsep}{0pt}
}{\end{itemize}}
\newenvironment{packed_enumerate}{
\begin{enumerate}
\setlength{\itemsep}{1pt}
\setlength{\parskip}{0pt}
\setlength{\parsep}{0pt}
\setlength{\headsep}{0pt}
\setlength{\topskip}{0pt}
\setlength{\topmargin}{0pt}
\setlength{\topsep}{0pt}
\setlength{\partopsep}{0pt}
}{\end{enumerate}}

\newcommand{\para}[1]{{\vspace{5pt} \bf \noindent #1 \hspace{8pt}}}
\newcommand{\tabincell}[2]{\begin{tabular}{@{}#1@{}}#2\end{tabular}}

\newcommand{\fixme}[1]{{#1}}

\newcommand\secspace{\vspace{0in}}

\begin{document}

%----------------copyright ----------------------

\CopyrightYear{2016} 
\setcopyright{acmcopyright}
\conferenceinfo{MobiSys'16,}{June 25-30, 2016, Singapore, Singapore}
\isbn{978-1-4503-4269-8/16/06}\acmPrice{\$15.00}
\doi{http://dx.doi.org/10.1145/2906388.2906420}
% --- End of Author Metadata ---
\clubpenalty=10000 
\widowpenalty = 10000

%---------------copyright ------------------------

\title{Defending against Sybil Devices in Crowdsourced Mapping Services}

\author{
\alignauthor Gang Wang$^\dag$, Bolun Wang$^\dag$, Tianyi Wang$^\dag$$^\ddag$,
  Ana Nika$^\dag$, Haitao Zheng$^\dag$, Ben Y. Zhao$^\dag$ \\
 \affaddr{$^\dag$Department of Computer Science, UC Santa Barbara} \\
 \affaddr{$^\ddag$Department of Electronic Engineering, Tsinghua University}\\
  \{gangw, bolunwang, tianyi, anika, htzheng, ravenben\}@cs.ucsb.edu
}

 % \affaddr{$^\dag$Department of Computer Science, UC Santa Barbara, Santa Barbara, CA, 93106} \\
 % \affaddr{$^\ddag$Department of Electronic Engineering, Tsinghua University, Beijing, China, 100084}

\maketitle

\begin{abstract}
Real-time crowdsourced maps such as Waze provide timely updates on traffic,
  congestion, accidents and points of interest.  In this paper, we
  demonstrate how lack of strong location authentication allows creation of
  software-based {\em Sybil devices} that expose crowdsourced map systems to a
  variety of security and 
  privacy attacks. Our experiments show that a single Sybil device with limited
  resources can cause havoc on Waze, reporting false congestion and accidents
  and automatically rerouting user traffic.  More importantly, we describe
  techniques to generate Sybil devices at scale, creating armies of virtual
  vehicles capable of remotely tracking 
  precise movements for large user populations while avoiding detection.  
  We propose a new approach to defend against Sybil devices based on {\em
    co-location edges}, authenticated records that attest to the one-time
  physical co-location of a pair of devices. Over time, co-location edges
  combine to form large {\em proximity graphs} that attest to
  physical interactions between devices, allowing scalable detection of
  virtual vehicles. 
  We demonstrate the efficacy of this approach using large-scale simulations, and
  discuss how they can be used to dramatically reduce the impact of attacks
  against crowdsourced mapping services.
 \end{abstract}

\secspace
\section{Introduction}
\secspace Crowdsourcing is indispensable as a real-time data gathering tool
for today's online services.  Take for example map and navigation services.
Both Google Maps and Waze use periodic GPS readings from mobile
devices to infer traffic speed and congestion levels on streets and highways.
Waze, the most popular crowdsourced map service, offers users more ways to
actively share information on accidents, police cars, and even contribute
content like editing roads, landmarks, and local fuel prices. This and the
ability to interact with nearby users made Waze extremely popular, with an
estimated 50 million users when it was acquired by Google for a reported
\$1.3 Billion USD in June 2013.  Today, Google integrates selected
crowdsourced data ({\em e.g.} accidents) from Waze into its own Maps
application.

Unfortunately, systems that rely on crowdsourced data are inherently vulnerable
to mischievous or malicious users seeking to disrupt or game the
system~\cite{Stefanovitch2014}. For example, business owners can badmouth
competitors by falsifying negative reviews on Yelp or TripAdvisor, and
FourSquare users can forge their physical locations for
discounts~\cite{mass12foursquare,foursq}. For location-based services, these
attacks are possible because there are no widely deployed tools to
authenticate the location of mobile devices.  In fact, there are few effective
tools today to identify whether the origin of traffic requests are real
mobile devices or software scripts.

The goal of our work is to explore the vulnerability of today's crowdsourced
mobile apps against {\em Sybil devices}, software scripts that appear to
application servers as ``virtual mobile devices.''\footnote{We refer to these
  scripts as Sybil devices, since they are the manifestations of Sybil
  attacks~\cite{Sybil} in the context of mobile networks.} While a single
Sybil device can damage mobile apps through misbehavior, larger groups of
Sybil devices can overwhelm normal users and significantly disrupt any
crowdsourced mobile app. In this paper, we identify techniques that allow
malicious attackers to reliably create large populations of Sybil devices
using software.  Using the context of the Waze crowdsourced map service, we
illustrate the powerful Sybil device attack, and then develop and evaluate
robust defenses against them.  

While our experiments and defenses are designed with Waze (and crowdsourced
maps) in mind, our results generalize to a wide range of mobile apps.  With
minimal modifications, our techniques can be applied to services ranging from
Foursquare and Yelp to Uber and YikYak, allowing attackers to cheaply emulate
numerous virtual devices with forged locations to overwhelm these systems via
misbehavior. Misbehavior can range from falsely obtaining coupons on
FourSquare/Yelp, gaming the new user coupon system in Uber, to imposing
censorship on YikYak. We believe our proposed defenses can be extended to
these services as well. We discuss broader implications of our work in
Section~\ref{sec:broad}.

\para{Sybil attacks in Waze.} In the context of Waze, our experiments reveal a
number of potential attacks by Sybil devices. First is simple {\em event
  forgery}, where devices can generate fake events to the Waze server,
including congestion, accidents or police activity that might affect user
routes.  Second, we describe techniques to reverse engineer mobile app APIs,
thus allowing attackers to create lightweight scripts that effectively
emulate a large number of virtual vehicles that collude under the control of
a single attacker. We call Sybil devices in Waze ``ghost riders.'' These
Sybils can effectively magnify the efficacy of any attack, and overwhelm
contributions from any legitimate users.  Finally, we discover a significant
privacy attack where ghost riders can silently and invisibly ``follow'' and
precisely track individual Waze users throughout their day, precisely mapping
out their movement to work, stores, hotels, gas station, and home.  We
experimentally confirmed the accuracy of this attack against our own
vehicles, quantifying the accuracy of the attack against GPS coordinates.
Magnified by an army of ghost riders, an attacker can potentially track the
constant whereabouts of millions of users, all without any risk of
detection.

\para{Defenses.} Prior proposals to address the location authentication problem have
limited appeal, because of reliance on widespread deployment of specialized
hardware, either as part of physical infrastructure, {\em i.e.}, cellular
base stations, or as modifications to mobile devices themselves.  Instead, we
propose a practical solution that limits the ability of Sybil devices to
amplify the potential damage incurred by any single attacker.  We introduce
{\em collocation edges}, authenticated records that attest to the one-time
physical proximity of a pair of mobile devices.  The creation of collocation
edges can be triggered opportunistically by the mapping service, {\em
  e.g.}, Waze.  Over time, collocation edges combine to form large {\em proximity
  graphs}, network structures that attest to physical interactions between
devices.  Since ghost riders cannot physically interact with real devices,
they cannot form direct edges with real devices, only indirectly through a
small number of real devices operated by the attacker.
Thus, the edges between an attacker and the rest of the network are limited
by the number of real physical devices she has, regardless of how many ghost
riders are under her control.  This reduces the problem of detecting ghost
riders to a community detection problem on the proximity
graph (The graph is seeded by a small number of trusted infrastructure
locations).

Our paper includes these key contributions:
\begin{packed_itemize}
\item We explore limits and impacts of single device attacks on
  Waze, {\em e.g.}, artificial congestion and events.
\item We describe techniques to create light-weight ghost riders, virtual
  vehicles emulated by client-side scripts, through reverse engineering of
  the Waze app's communication protocol with the server.
\item We identify a new privacy attack that allows ghost riders to virtually
  follow and track individual Waze users in real-time, and describe
  techniques to produce precise, robust location updates.  
\item We propose and evaluate defenses against ghost riders, using {\em
    proximity graphs} constructed with edges representing authenticated
  collocation events between pairs of devices.  Since collocation can only
  occur between pairs of physical devices, proximity graphs limit the number
  of edges between real devices and ghost riders, thus isolating groups of
  ghost riders and making them detectable using community detection
  algorithms.
\end{packed_itemize}

\secspace
\section{Waze Background}
\label{sec:back}
\secspace 

Waze is the most popular crowdsourced navigation app on smartphones, with
more than 50 million users when it was acquired by Google in June
2013~\cite{waze-google3}.  Waze collects GPS values of users' devices to
estimate real-time traffic. It also allows users to report on-road events
such as accidents, road closures and police vehicles, as well as curating
points of interest, editing roads, and even updating local fuel prices.  Some
features, {\em e.g.}, user reported accidents, have been integrated into
Google Maps~\cite{waze-google1}. Here, we briefly describe the key
functionality in Waze as context for our work.

\para{Trip Navigation.} Waze's main feature is assist users to find
the best route to their destination and turn-by-turn navigation. 
Waze generates aggregated real-time traffic updates using GPS data from its
users, and optimizes user routes both during trip planning and during
navigation. If and when traffic congestions is detected, Waze automatically
re-routes users towards an alternative. 

\para{Crowdsourced User Reports.} Waze users can generate real-time {\em
  event reports} on their routes to inform others about ongoing
incidents. Events range from accidents to road closures, hazards, and even
police speed traps.  Each report can include a short note with a
photo. The event shows up on the map of users driving towards the
reported location. As users get close, Waze pops up a window to let the user
``say thanks,'' or report the event is ``not there.''  If multiple
users choose ``not there'', the event will be removed.
Waze also merges multiple reports of the same event type at the same
location into a single event.

\para{Social Function.} To increase user engagement, Waze supports simple
social interactions. Users can see avatars and locations of nearby users.
Clicking on a user's avatar shows more detailed user information, including
nickname, ranking, and traveling speed. Also, users can send messages
and chat with nearby users. This social
function gives users the sense of a large community. Users can elevate their
rankings in the community by contributing and receiving ``thanks'' from others.

\begin{figure}[t]
 \centering
\begin{minipage}{0.44\textwidth}
        \centering
  \includegraphics[width=1\textwidth]{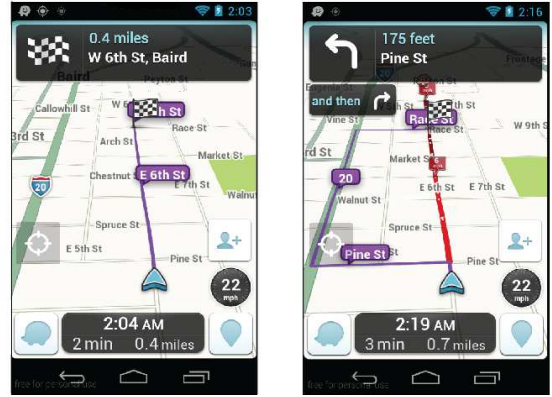}
  \vspace{-0.1in}
\end{minipage}
  \caption{Before the attack (left), Waze shows the fastest route for
          the user. After the attack (right), the user gets
          automatically re-routed by the fake traffic jam.}
  \label{fig:screen}
\vspace{-0.05in}
\end{figure}

\begin{figure*}
\centering
    \subfigure[Highway]{
      \includegraphics[width=0.31\textwidth]{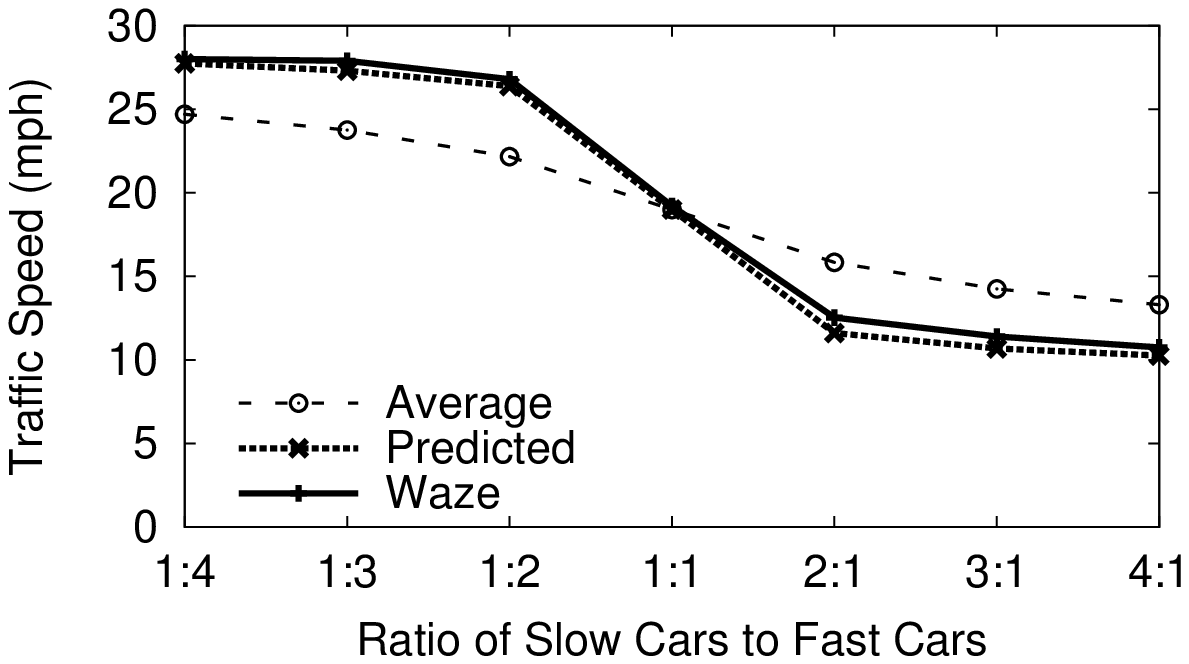}
      % \caption{Speed function highway \fixme{need a better one}.}
      \vspace{-0.05in}
      \label{fig:speed_function_ss_highway}
      \vspace{-0.05in}
    }
    \hfill
    % \hspace{0.1in}
    \subfigure[Local Road]{
      \includegraphics[width=0.31\textwidth]{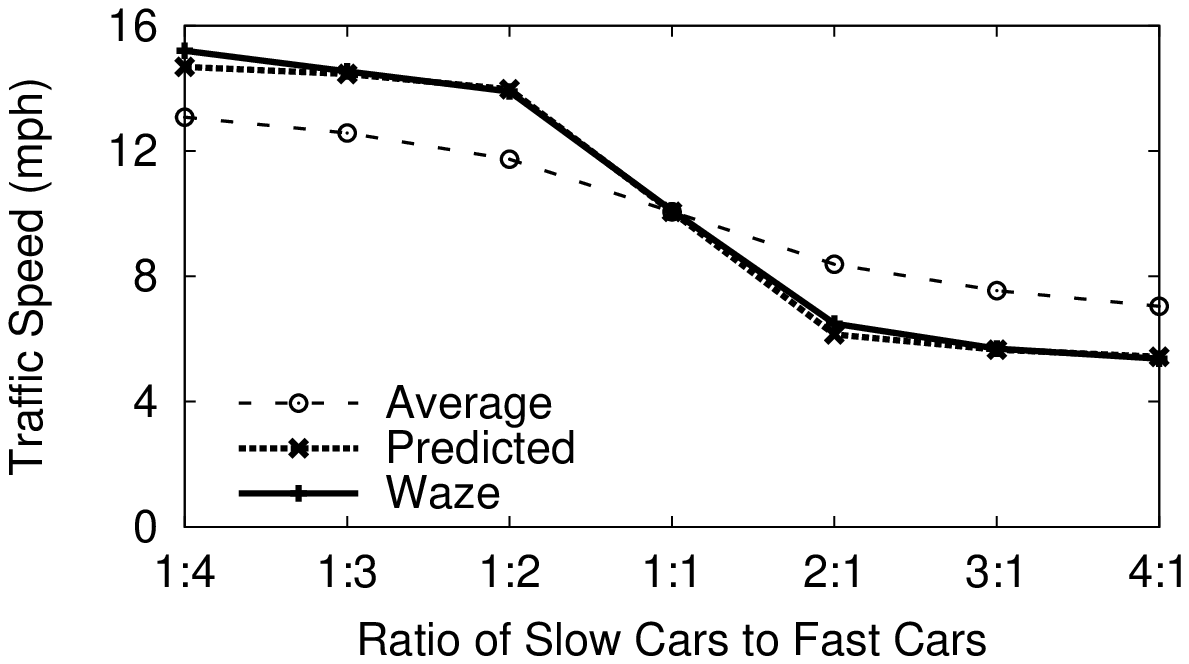}
      % \caption{Speed function local \fixme{need a better one}.}
     \vspace{-0.05in}
      \label{fig:speed_function_ss_local}
      \vspace{-0.05in}
    }
    \hfill
    % \hspace{0.1in}
    \subfigure[Residential]{
      \includegraphics[width=0.31\textwidth]{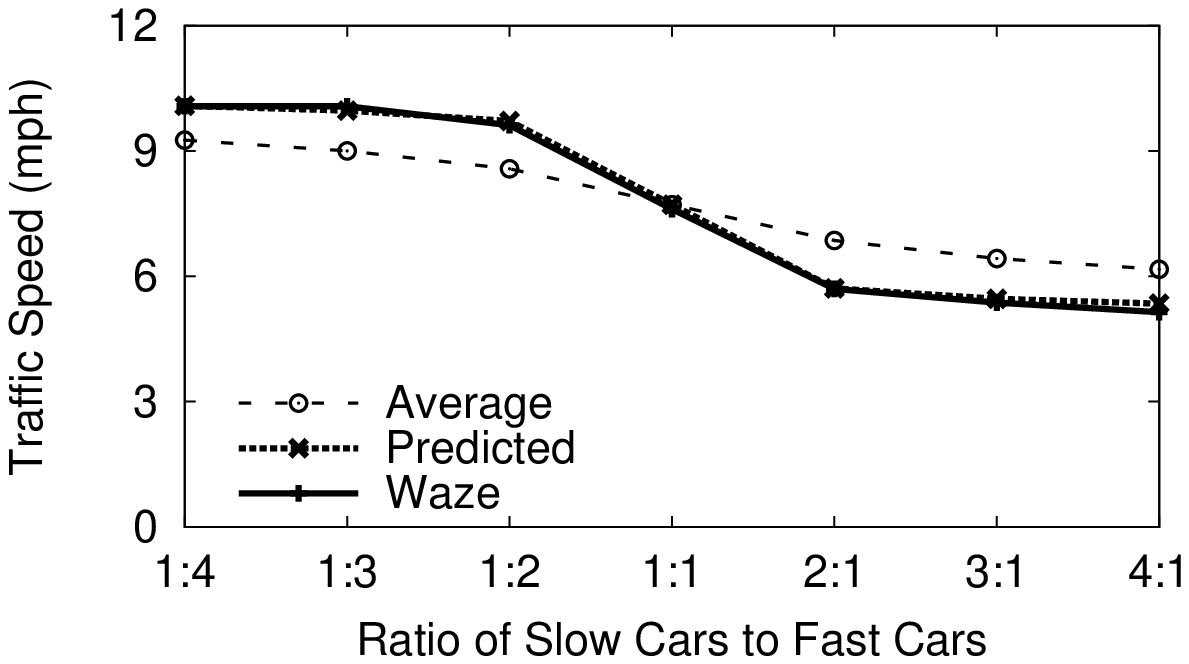}
     \vspace{-0.05in} 
      \label{fig:speed_function_ss_residential}
      \vspace{-0.05in}
    }
\vspace{-0.05in}
  \caption{The traffic speed of the road with respect to different combinations of
    number of slow cars and fast cars. We show that  Waze is not using
    the average speed of all cars, and our inferred function can
    correctly predict the traffic speed displayed on Waze. 
  }
\vspace{-0.05in}
  \label{fig:speed_ss}
\end{figure*}

\section{Attacking Crowdsourced Maps} 
\label{sec:traffic}
\secspace
In this section, we describe basic attacks to manipulate Waze
by generating false road events and fake traffic
congestion. Since Waze relies on real-time data for trip planning
and route selection, these attacks can influence
user's routing decisions. Attackers can attack specific users by forging
congestion to force automatic rerouting on their trips. 
The attack is possible because Waze has no reliable authentication on user
reported data, such as their device GPS.

We first discuss experimental ethics and steps we took to limit impact on
real users. Then, we describe basic mechanisms and resources needed to launch
attacks, and use controlled experiments on two attacks to understand their
feasibility and limits. One attack creates fake road events at
arbitrary locations, and the other seeks to generate artificial traffic hotspots to
influence user routing. 

\secspace
\subsection{Ethics}
\secspace Our experiments seek to understand the feasibility and limits of
practical attacks on crowdsourcing maps like Waze. We are very aware of the
potential impact to real Waze users from any experiments. We consulted our
local IRB and have taken all possible precautions to ensure that our
experiments do not negatively impact real Waze users. In particular, we
choose experiment locations where user population density is extremely low
(unoccupied roads), and only perform experiments at low-traffic hours, {\em
  e.g.}, between 2am and 5am. During the experiments, we continuously scan
the entire experiment region and neighboring areas, to ensure no other Waze
users (except our own accounts) are within miles of the test area. If any
Waze users are detected, we immediately terminate all running experiments. Our study
received the IRB approval under protocol\# COMS-ZH-YA-010-7N. 

Our work is further motivated by our view of the risks of inaction versus
risks posed to users by our study. On one hand, we can and have minimized
risk to Waze users during our study, and we believe our experiments have not
affected any Waze users. On the other hand, we believe the risk to millions
of Waze users from pervasive location tracking (described in
Section~\ref{sec:track}) is realistic and potentially very damaging. We feel
that investigating these attacks and identifying these risks to the broad
community at large was the ethically correct course of action.  Furthermore,
full understanding of the attacks was necessary to design an effective and
{\em practical} defense.  Please see Appendix A for more detailed information
on our IRB approval and steps taken towards responsible disclosure.

\subsection{Basic Attack: Generating Fake Events}
Launching attacks against crowdsourced maps like Waze requires
three steps: automate input to mobile devices that run the Waze app; control
the device GPS and simulate device movements ({\em e.g.}, car driving); obtain
access to {\em multiple} devices. All three are easily achieved using
widely available mobile device emulators.

Most mobile emulators run a full OS ({\em e.g.}, Android, iOS) down to the
kernel level, and simulate hardware features such as camera, SDCard and GPS.
We choose the GenyMotion Android emulator~\cite{genymotion} for its
performance and reliability.  Attackers can automatically control the
GenyMotion emulator via Monkeyrunner scripts~\cite{monkeyrunner}.  They can
generate user actions such as clicking buttons and typing text, and feed
pre-designed GPS sequences to the emulator (through a command line interface)
to simulate location positioning and device movement.  By controlling the
timing of the GPS updates, they can simulate any ``movement speed'' of the
simulated devices.

Using these tools, attackers can generate fake events (or alerts) at a given
location by setting fake GPS on their virtual devices.  This includes any
events supported by Waze, including accidents, police, hazards, and road
closures. We find that a single emulator can generate any event at arbitrary
locations on the map. We validate this using experiments on a variety of
unoccupied roads, including highways, local and rural roads
(50+ locations, 3 repeated tests each). Note that our experiments only involve data in the
  Waze system, and do not affect real road vehicles not running the Waze app.
  Thus ``unoccupied'' means no vehicles on the road with mobile devices
  actively running the Waze app. After creation, the fake event stays
on the map for about 30 minutes. Any Waze user can report that an
event was ``not there.''  We find it takes two consecutive ``not theres'' (without any ``thanks''
in between) to delete the event.  Thus an attacker can ensure an event
persists by occasionally ``driving'' other virtual devices to the region and
``thanking'' the original attacker for the event report.

\secspace
\subsection{Congestion and Traffic Routing}
\label{sec:traffic2}
\secspace

A more serious attack targets Waze's real-time trip routing function. Since
route selection in Waze relies on predicted trip time, attackers can
influence routes by creating ``fake'' traffic hotspots at specific
locations. This can be done by configuring a group of virtual vehicles to
travel slowly on a chosen road segment.

We use controlled experiments to answer two questions. First, under what
conditions can attackers successfully create traffic hotspots? Second, how
long can an artificial traffic hotspot last?  We select three low-traffic
roads in the state of Texas that are representative of three popular road
types based on their speed limit---Highway (65 mph), Local (45 mph) and
Residential (25 mph).  To avoid real users, we choose roads in low population
rural areas, and run tests at hours with the lowest traffic volumes (usually
3-5AM).  We constantly scan for real users in or nearby the experimental
region, and reset/terminate experiments if users come close to an area with
ongoing experiments. Across all our experiments, only 2 tests were
terminated due to detected presence of real users nearby. 
Finally, we % perform
% spot sanity checks with
 have examined different road types and hours of the day to ensure
they do not introduce bias into our results.

\para{Creating Traffic Hotspots.}
Our experiment shows that it only takes one slow moving car to
create a traffic congestion, when there are no real Waze users
around. 
% our virtual vehicle provides the only input to Waze. 
Waze displays a red overlay on the road to indicate traffic
congestion (Figure~\ref{fig:screen}, right). Different road
types have different congestion thresholds, with thresholds strongly
correlated to the speed
limit.  % As shown in Figure~\ref{fig:speed_threshold}, congestion
The congestion thresholds for Highway, Local and Residential
roads are 40mph, 20mph and 15mph, respectively.

To understand if this is generalizable, we repeat our tests on other
unoccupied roads in different states and countries. We picked 18 roads in
five states in the US (CO, MO, NM, UT, MS) and British Columbia, Canada. In
each region, we select three roads with different speed limits (highway,
local and residential). We find consistent results: a single virtual vehicle can
always generate a traffic hotspot; and the congestion thresholds were
consistent across different roads of the same speed limit.

\para{Outvoting Real Users.} Generating traffic hotspot in practical
scenarios faces a challenge from real Waze users who drive at normal
(non-congested) speeds: attacker's virtual vehicles must ``convince'' the
server there's a stream of slow speed traffic on the road even as real
users tell the server otherwise. We need to understand how Waze aggregated
multiple inputs to estimate traffic speed.

We perform an experiment to infer this aggregation function used by Waze. We
create two groups of virtual vehicles: $N_s$ slow-driving cars with speed
$S_s$, and $N_f$ fast-driving cars with speed $S_f$; and they all pass the
target location at the same time. We study the congestion reported by Waze to
infer the aggregation function.  Note that the server-estimated traffic speed
is visible on the map {\em only if} we formed a traffic hotspot.
%To make sure the traffic hotspot is formed, we 
We achieve this by setting the speed tuple
($S_s$, $S_f$) to (10mph, 30mph) for Highway, (5, 15) for Local and (5, 10)
for Residential. 
% and vary $N_s$ and $N_f$.

As shown in Figure~\ref{fig:speed_ss}, when we vary the ratio of slow cars
over fast cars ($N_s$:$N_f$), the Waze server produces different final
traffic speeds.  We observe that Waze does not simply compute
an ``average'' speed over all the cars. Instead, it uses a weighted average
with higher weight on the majority cars' speed. We infer an aggregation
function as follows:
\begin{equation*} 
 %  S_{waze}= \frac{S_{max} N_{max}+S_{avg}N_{min}}{N_s+N_f}
    S_{waze}= \frac{S_{max}\cdot max(N_s,N_f)+S_{avg}\cdot min(N_s, N_f)}{N_s+N_f}
\end{equation*}
%$N_{max} = max(N_s, N_f)$, $N_{min} = min(N_s, N_f)$, 
where $S_{avg} = \frac{S_sN_s+S_fN_f}{N_s+N_f}$, and $S_{max}$ is the speed of the group with $N_{max}$ cars. 
%Using this function, the final speed leans towards the speed of the
%bigger group. 
As shown in Figure~\ref{fig:speed_ss}, our function can predict Waze's
aggregate traffic speed accurately, for all different types of roads
in our test.
For validation purposes, we run another set of experiments by raising
$S_f$ above the hotspot thresholds (65mph, 30mph and 20mph
respectively for the three roads).  We can still form traffic hotspots by using more
slow-driving cars ($N_s > N_f$), and our function can still predict the
traffic speed on Waze accurately.

\begin{figure}
  \centering
\begin{minipage}[t]{0.43\textwidth}
  \includegraphics[width=1\textwidth]{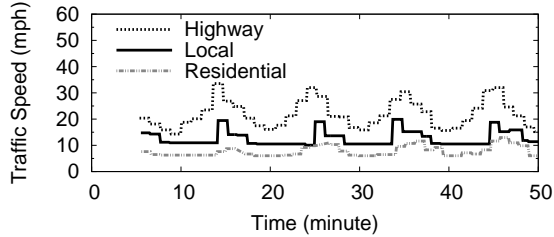}\\
\end{minipage}
\vspace{-0.05in}
  \caption{\small Long-last traffic jam created by slow cars
    driving-by.}\label{fig:repeat_driving}
\vspace{-0.05in}
\end{figure}

\para{Long-Lasting Traffic Congestion.} A traffic hotspot will last
for 25-30 minutes if no other cars drive by. Once aggregate speed
normalizes, the congestion event is
dismissed within 2-5 minutes. To create a long-lasting virtual
traffic jam, attackers can simply keep sending slow-driving cars to the
congestion area to resist the input from real users. We validate this using a simple, 50-minute long
experiment where 3 virtual vehicles create a persistent congestion by
driving slowly through an area, and then looping back every 10
minutes. Meanwhile, 2 other virtual cars emulate legitimate drivers that
pass by at high speed every 10 minutes. As shown in
Figure~\ref{fig:repeat_driving}, the traffic hotspot persists for
the entire experiment period.

\para{Impact on End Users.} Waze uses real-time traffic data to optimize
routes during trip planning.  Waze estimates the end-to-end trip time and
recommends the fastest route.  Once on the road, Waze continuously estimates
the travel time, and automatically reroutes if the current route becomes
congested.  An attacker can launch physical attacks by placing fake traffic
hotspots on the user's original route. While congestion alone does not
trigger rerouting, Waze reroutes the user to a detour when the estimated
travel time through the detour is shorter than the current congested route
(see Figure~\ref{fig:screen}). % Note that when users are driving,

We also note that Waze data is used by Google Maps, and therefore can
potentially impact their 1+ billion users~\cite{googlemap4}.  Our experiment
shows that artificial congestion do not appear on Google Maps, but fake
events generated on Waze are displayed on Google Maps without verification,
including ``accidents'', ``construction'' and ``objects on road''.  Finally,
event updates are synchronized on both services, with a 2-minute delay and
persist for a similar period of time ({\em e.g.}, 30 minutes).

\begin{figure}[t]
 \centering
% \begin{minipage}{0.42\textwidth}
        \centering
	\includegraphics[width=0.35\textwidth]{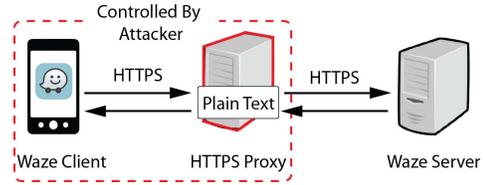}
	 % \vspace{-0.05in}
	\caption{Using a HTTPS proxy as man-in-the-middle
          to intercept traffic between Waze client and server. }
	% \caption{To reverse engineer Waze APIs, attacker uses a HTTPS
        %   proxy as man-in-the-middle to intercept traffic between
        %   Waze client and server. Attacker only needs to pre-install
        %   the proxy's root certificate to the phone. }
	\label{fig:https}
% \end{minipage}
% \vspace{-0.05in}
\end{figure}

\secspace
\section{Sybil Attacks} 
\label{sec:scale}
\secspace 
So far, we have shown that attackers using emulators can create
``virtual vehicles'' that manipulate the Waze map. An attacker can generate
much higher impact using a large group of virtual vehicles (or {\em
  Sybils}~\cite{Sybil}) under control.  In this section, we describe
techniques to produce light-weight virtual vehicles in Waze, and explore the
scalability of the group-based attacks.  We refer to large groups of virtual
vehicles as ``ghost riders'' for two reasons.  First, they are easy to create
en masse, and can travel in packs to outvote real users to generate more
complex events, {\em e.g.}, persistent traffic congestion.  Second, as
we show in \S\ref{sec:track}, they can make themselves invisible to nearby
vehicles.

\para{Factors Limiting Sybil Creation.}
We start by looking at the limits of the large-scale Sybil attacks on Waze.
First, we note user accounts do not pose a challenge to attackers, since
account registration can be fully automated. We found that a single-threaded
Monkeyrunner script could automatically register 1000 new accounts in a
day. Even though the latest version of Waze app requires SMS
  verification to register accounts, attackers can use older versions of APIs
  to create accounts without verification. Alternatively, accounts can
    be verified through disposable phone/SMS
    services~\cite{phone_ccs14}.

The limiting factor is the scalability of vehicle emulation.  Even though
emulators like GenyMotion are relatively lightweight, each instance still
takes significant computational resources. For example, a MacBookPro with 8G
of RAM supports only 10 simultaneous emulator instances. For this, we explore
a more scalable approach to client emulation that can increase the number of
supported virtual vehicles by orders of magnitude. Specifically, we reverse
engineer the communication APIs used by the app, and replace emulators with
simple Python scripts that mimic API calls.

\para{Reverse Engineering Waze APIs.}
The Waze app uses HTTPS to communicate with the server, so API details cannot
be directly observed by capturing network traffic (TLS/SSL
encrypted). However, an attacker can still intercept HTTPS traffic, by
setting up a proxy~\cite{charles} between her phone and Waze server as a
man-in-the-middle attack~\cite{SMVHunters:NDSS14, sslsp:2014}.  As shown in
Figure~\ref{fig:https}, an attacker needs to pre-install the proxy server's
root Certificate Authorities (CA) to her own phone as a ``trusted CA.''  This
allows the proxy to present self-signed certificates to the phone claiming to
be the Waze server. The Waze app on the phone will trust the proxy (since
the certificate is signed by a ``trusted CA''), and establish HTTPS connections
with the proxy using proxy's public key. On the proxy side, the attacker can
decrypt the traffic using proxy's private key, and then forward traffic
from the phone to Waze server through a separate TLS/SSL channel. The proxy
then observes traffic to the Waze servers and extracts the API
calls from plain text traffic.

Hiding API calls using traffic encryption is fundamentally challenging,
because the attacker has control over most of the components in the
communication process, including phone, the app binary, and the proxy. A
known countermeasure is certificate pinning~\cite{SSLpinning2013}, which
embeds a copy of the server certificate within the app.  When the app makes
HTTPS requests, it validates the server-provided certificate with its known
copy before establishing connections. However, dedicated attackers can
extract and replace the embedded certificate by disassembling the app binary
or attaching the app to a debugger~\cite{blackhat12ssl,Decomple2011}.

\para{Scalability of Ghost Riders.} 
With the knowledge of Waze APIs, we build extremely lightweight Waze clients
using python scripts, allocating one thread for each client. Within
each thread, we log in to the app using a separate account, and maintain a
live session by sending periodic GPS coordinates to the Waze server.
The Python client is a full Waze client, and can report fake events using the API.
Scripted emulation is highly scalable. We run 1000
virtual vehicles on a single Linux Dell PowerEdge Server (Quad Core, 2GB
RAM), and find that at steady state, 1000 virtual devices only
introduces a small overhead: 11\% of memory usage, 2\% of CPU and
420 Kbps bandwidth. In practice, attackers can easily run
tens of thousands of virtual devices on a commodity server. 

Finally, we experimentally confirm the practical efficacy and scalability of
ghost riders.  We chose a secluded highway in rural Texas, and used 1000
virtual vehicles (hosted on a single server and single IP) to generate a
highly congested traffic hotspot. We perform our experiment in the middle of
the night after repeated scans showed no Waze users within miles of our test
area.  We positioned 1000 ghost riders one after another, and drove them
slowly at 15 mph along the highway, looping them back every 15 minutes for an
entire hour.  The congestion shows up on Waze 5 minutes after our test began,
and stayed on the map during the entire test period.  No problems were
observed during our test, and tests to generate fake events (accidents etc.)
also succeeded.

\secspace
\section{User Tracking Attack} 
\label{sec:track}
\secspace 
Next, we describe a powerful new attack on user privacy, where
virtual vehicles can track Waze users continuously without risking detection
themselves.  By exploiting a key social functionality in Waze, attackers can
remotely follow (or stalk) any individual user in real time.  This is
possible with single device emulation, but greatly amplified with the help of
large groups of ghost riders, possibly tracking large user populations
simultaneously and putting user (location) privacy at great risk.  We start
by examining the feasibility (and key enablers) of this attack. We then
present a simple but highly effective tracking algorithm that follows
individual users in real time, which we have validated using real life
experiments (with ourselves as the targets).

The only way for Waze users to avoid tracking is to go ``invisible'' in Waze.
However, doing so forfeits the ability to generate reports or message other
users. Users are also reset to ``visible''  each time the Waze app opens.

\secspace
\subsection{Feasibility of User Tracking}
\secspace 
A key feature in Waze allows users to socialize with others on the
road. Each user sees on her screen icons representing the locations of
nearby users, and can chat or message with them through the app. Leveraging
this feature, an attacker can pinpoint any target who has the Waze app
running on her phone. By constantly ``refreshing'' the app screen (issuing an
update query to the server), an attacker can query the victim's GPS location
from Waze in real time.
% \footnote{Waze user reports her GPS to the server every 2 minutes.}.
To understand this capability, we perform detailed measurements on Waze to
evaluate the efficiency and precision of user tracking. % This includes the

\para{Tracking via User Queries.} A Waze client periodically requests updates
in her nearby area, by issuing an update query with its GPS coordinates and a
rectangular ``search area.'' This search area can be set to any location on
the map, and does not depend on the requester's own location. The server
returns a list of users located in the area, including userID, nickname,
account creation time, GPS coordinates and the GPS timestamp. Thus an
attacker can find and ``follow'' a target user by first locating them at any
given location (work, home) and then continuously following them by issuing
update queries centered on the target vehicle location, all automated by scripts.

\begin{table*}[t]
\centering{
% \resizebox{0.85\textwidth}{!}{
% \small{
\begin{tabular}{|c|c|c||c|c|c|c|c|}
\hline
 	 Location  & \tabincell{c}{Route \\ Length (Mile)} & 
         \tabincell{c}{Travel \\ Time (Minute)}  &
         \tabincell{c}{ GPS Sent \\ By Victim}	&
	 \tabincell{c}{ GPS Captured \\ by Attacker}	&
	 \tabincell{c}{Followed to \\Destination? }	&
         \tabincell{c}{Avg. Track \\ Delay (Second)} &
          \tabincell{c}{Waze User Density \\ (\# of Users / mile$^2$)}\\
\hline
	City A 	& 12.8 	&  35 & 18 & 16 & Yes &  43.79 & 56.6\\
\hline
	Highway B & 36.6 & 40  & 20 & 19 & Yes &9.24 & 2.8 \\	
\hline
\end{tabular}}
% }
% \vspace{-0.05in}
\caption{Tracking Experiment Results. }
\label{tab:track_detail}
% \vspace{-0.05in}
\end{table*}

\begin{figure}[t]
 \centering
	\includegraphics[width=0.4\textwidth]{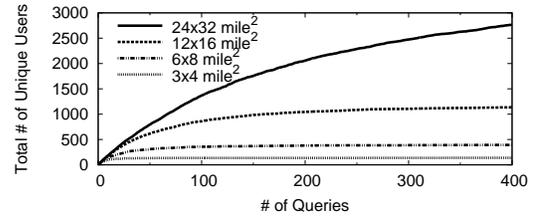}
	% \vspace{-0.05in}
	\caption{\# of queries vs. unique returned users
          in the area.}
	\label{fig:down1}
\vspace{-0.1in}
\end{figure}

\para{Overcoming Downsampling.} 
The user query approach faces a downsampling challenge, because Waze responds
to each query with an ``incomplete'' set of users, {\em i.e.}, up to 20 users
per query regardless of the search area size.  This downsampled result is
necessary to prevent flooding the app screen with too many user icons, but it
also limits an attacker's ability to follow a moving target.

This downsampling can be overcome by simply repeatedly querying the system
until the target is found. 
We perform query measurements on four test areas (of different sizes
between $3\times4$ mile$^2$ and $24\times32$ mile$^2$) in the downtown area
of Los Angeles (City A, with 10 million residents as of 2015).
For each area, we issue 400 queries within 10 seconds, and examine the number
of unique users returned by all the queries. 
% Since Waze applies a rate limit of 2 queries per second per-account, we use
% 20 accounts to perform this experiment.
Results in Figure~\ref{fig:down1} show that the number of unique users
reported converges after 150-250 queries for the three small search areas
($\leq 12\times16$ mile$^2$).  For the area of size 24$\times$32 mile$^2$,
more than 400 queries are required to reach convergence.

\begin{figure}[t]
\centering
	\includegraphics[width=0.4\textwidth]{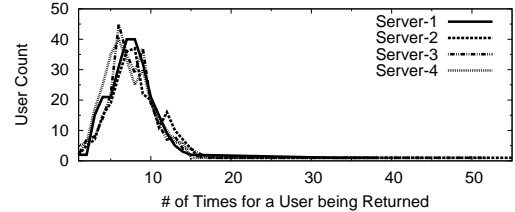}
	% \vspace{-0.05in}
	\caption{User's number of appearances in the returned
          results  ($6 \times 8$ mile$^2$ area).}
	\label{fig:down3}
% \vspace{-0.05in}
\end{figure}

\fixme{
We confirm this ``downsampling'' is uniformly random, by comparing our
measurement results to a mathematical model that projects the statistics of
query results assuming uniform-random sampling.  Consider total $M$
users in the search area. The probability of a user $x$
getting sampled in a single round of query (20 users per query) is
$P(x)=\frac{20}{M}$. Over $N$ queries, the number of appearances per
user should follow a Binomial Distribution~\cite{Binomial:1988} with mean
$N\cdot\frac{20}{M}$. Figure~\ref{fig:down3} plots the measured user
appearances for the four servers on the $6 \times 8$ mile$^2$ area with
$N=100$. The measured statistics follow the projected Binomial Distribution
(the measured mean values closely match the theoretical expectation).  This
confirms that the downsampling is indeed random, and thus an attacker can
recover a (near) complete set of Waze users with repeated queries. 
While the number of queries required increases superlinearly with area
size, a complementary technique is to divide an area into smaller,
fixed size partitions and query each partition's users in parallel.

We also observe that user lists returned by different Waze servers had only
a partial overlap (roughly 20\% of users from each server were unique to that
server).  This ``inconsistency'' across servers is caused by 
synchronization delay among the servers.  Each user only sends its GPS
coordinates to a single server which takes 2-5 minutes to 
% and it takes 2-5 minutes for this update to
propagate to other servers. Therefore, a complete user set requires queries
to cover all Waze servers.  At the time of our experiments, the number
of Waze servers could be traced through app traffic and could be covered by a
moderate number of querying accounts.
}

\para{Tracking Users over Time.} Our analysis found that each active Waze app
updates its GPS coordinates to the server every 2 minutes, regardless of
whether the user is mobile or stationary.  Even when running in the
background, the Waze app reports GPS values every 5 minutes.  As long as the
Waze app is open (even running in the background), the user's location is
continuously reported to Waze and potential attackers. Clearly, a more
conservative approach to managing location data would be extremely helpful
here.

We note that attackers can perform long-term tracking on a target user ({\em
  e.g.}, over months). The attacker needs a persistent ID associated to the
target. The ``userID'' field in the metadata is insufficient, because it is a
random ``session'' ID assigned upon user login and is released when the user
kills the app. However, the ``account creation time'' can serve as a
persistent ID, because a) it remains the same across the user's different
login sessions, and b) it is precise down to the second, and is sufficiently
to uniquely identify single users in the same geographic area.
% This allows the attacker to persistently track a target user. 
While Waze can remove the ``account creation time'' field from metadata, 
a persistent attacker can overcome this by analyzing the victim's mobility
pattern. For example, the attacker can identify a set of locations where the
victim has visited frequently or stayed during the past session, mapping to
home or workplace. Then the attacker can assign a ghost rider to constantly
monitor those areas, and re-identify the target once her icon shows
up in a monitored location, {\em e.g.}, home.

\para{Stealth Mode.} We note that attackers remain 
invisible to their targets, because queries on any specific geographic
area can be done by Sybils operating ``remotely,'' i.e. claiming to be in a
different city, state or country.  Attackers can enable their
``invisible'' option to hide from other nearby users.
Finally, disabling these features still does not make the
attacker visible.  Waze only updates each user's ``nearby'' screen every 2
minutes (while sending its own GPS update to the servers).  Thus a tracker
can ``pop into'' the target's region, query for the target, and then move out
of the target's observable range, all before the target can update and detect
it.

\begin{figure}[t]
\centering
\begin{minipage}{0.48\textwidth}
	\includegraphics[width=1\textwidth]{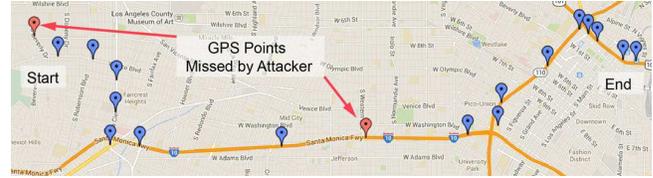}
	% \vspace{-0.05in}
	\caption{A graphical view of the tracking result in Los
          Angeles downtown (City A). Blue dots
          are GPS points captured by the attacker and the red dots are those
          missed by the attacker.}
	\label{fig:routeLA}
\end{minipage}
% \vspace{-0.05in}
\end{figure}

\secspace
\subsection{Real-time Individual User Tracking}
\label{sec:trackuser}
\secspace 

To build a detailed trace of a target user's movements, an attacker
first bootstraps by identifying the target's icon on the map. This can be
done by identifying the target's icon while confirming her physical presence
at a time and location. The attacker centers its search area on the victim's
location, and issues a large number of queries (using Sybil accounts) until
it captures the next GPS report from the target. If the target is moving, the
attacker moves the search area along the target's direction of movement and
repeats the process to get updates. 

\para{Experiments.} To evaluate its effectiveness, we performed experiments
by tracking one of our own Android smartphones and one of our virtual
devices. Tracking was effective in both cases, but we experimented more with
tracking our virtual device, since we could have it travel to any location.
Using the OSRM tool~\cite{osrm}, we generate detailed GPS traces of two
driving trips, one in downtown area of \fixme{Los Angeles} (City A), and 
% a metropolitan city (City A), and 
one along the interstate \fixme{highway-101} (Highway B). The target device
uses a realistic driving speed based on average traffic speeds estimated by Google
Maps during the experiment. The attacker used 20 virtual devices
to query Waze simultaneously in a rectangular search area of size
$6\times 8$ mile$^2$. \fixme{
This should be sufficient to track
the GPS update of a fast-driving car (up to 160 mph). } 
Both experiments were during morning hours, and we logged both the network
traffic of the target phone and query data retrieved by the attacker. Note
that we did not generate any ``events'' or otherwise affect the Waze system
in this experiment.

\para{Results.}  Table~\ref{tab:track_detail} lists the results of tracking
our virtual device, and Figure~\ref{fig:routeLA} presents a graphical
view of the City A result.  For both routes, the
attacker can consistently follow the victim to her destination, though the
attacker fails to capture 1-2 GPS points out of the 18-20 reported. 
For City A, the tracking delay, {\em i.e.}, the time spent to
capture the subsequent GPS of the victim, is larger
(averaging 43s rather than 9s).  This is because the downtown area has a
higher Waze user density, and required more rounds of
queries to locate the target.

Our experiments represent two highly challenging
({\em i.e.}, worst case) scenarios for the attacker. The high density of Waze
users in City A downtown is makes it challenging to
locate a target in real time with downsampling. On Highway B,
the target travels at a high speed ($\sim$60mph), putting a stringent time
limit on the tracking latency, {\em i.e.}, the attacker must capture the
target before he leaves the search area.  The success of both experiments
confirms the effectiveness and practicality of the proposed attack.

\section{Defenses} 
\label{sec:defense}
\secspace

In this section, we propose defense mechanisms to significantly limit the
magnitude and impact of these attacks. 
While individual devices can inflict limited damage, an attacker's
ability to control a large number of virtual vehicles
at low cost elevates the severity of the attack in both quantity and
quality. Our priority, then, is to restrict the number of
ghost riders available to each attacker, thus increasing the cost per
``vehicle'' and reducing potential damage.

The most intuitive approach is perform strong location authentication, so
that attackers must use real devices physically located at the actual
locations reported. This would make ghost riders as expensive to operate as
real devices.  Unfortunately, existing methods for location authentication do
not extend well to our context. Some proposals solely rely on trusted
infrastructures ({\em e.g.}, wireless access points) to verify the physical
presence of devices in close
proximity~\cite{hotmobile10locpro1,hotmobile09locproof}.  However, this
requires large scale retrofitting of cellular celltowers or installation of
new hardware, neither of which is practical at large geographic scales.
Others propose to embed tamperproof location hardware on mobile
devices~\cite{ndss14pay,hotmobile10sensing}, which incurs high cost per user,
and is only effective if enforced across all devices. For our purposes, we
need a scalable approach that works with current hardware, without incurring
costs on mobile users or the map service (Waze).

\begin{figure*}[t]
\begin{minipage}{0.48\textwidth}
 \centering
\includegraphics[width=1\textwidth]{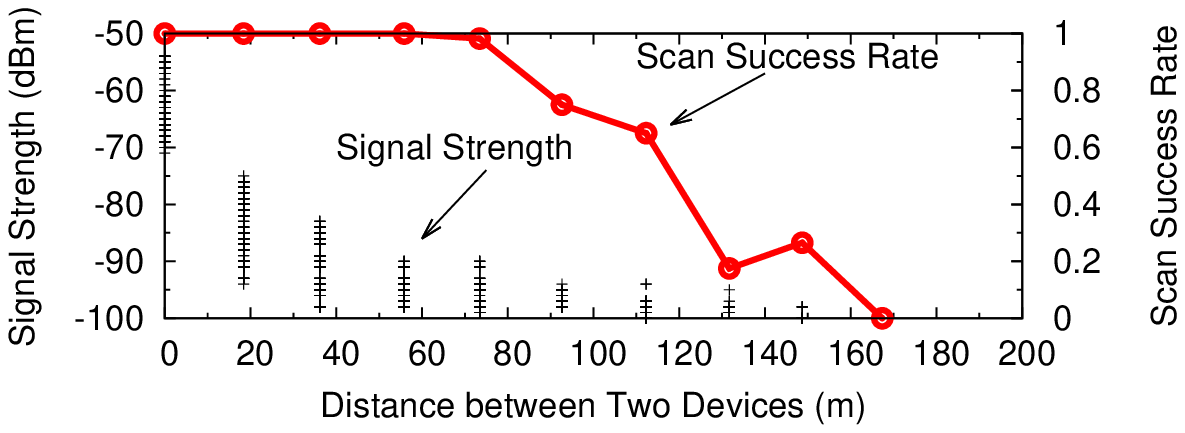}
\vspace{-0.05in}	
\caption{WiFi signal strength and scan success rate with respect to car distance in
          static scenarios.}
	\label{fig:static1}
\end{minipage}
\vspace{-0.01in}
\hfill 
\begin{minipage}{0.48\textwidth}
 \centering
	\includegraphics[width=1\textwidth]{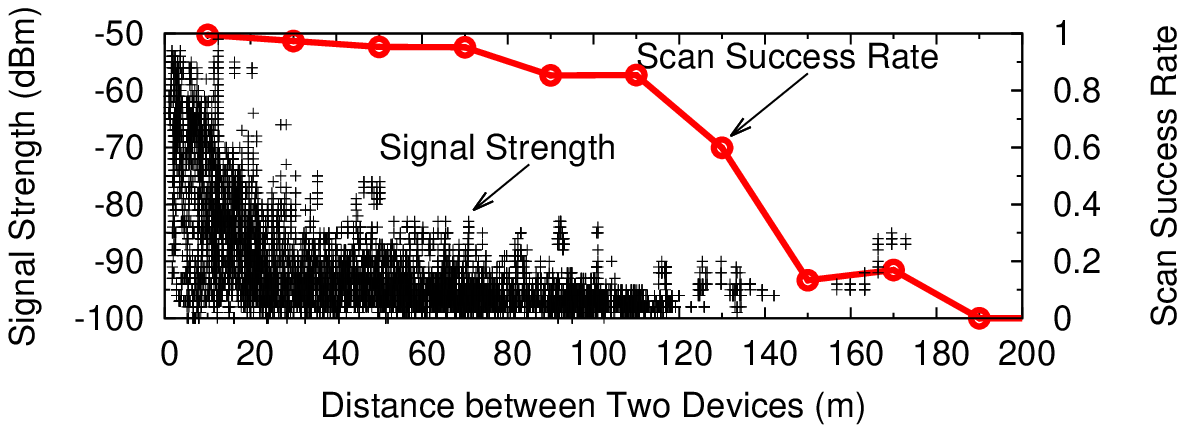}
\vspace{-0.05in}
	\caption{WiFi signal strength and scan success rate with respect to car distance in
          driving scenarios.}
	\label{fig:drive2}
\end{minipage}
\vspace{-0.01in}
\end{figure*}

\secspace
\subsection{Sybil Detection via Proximity Graph}  
\secspace
Instead of optimizing per-device location authentication, 
our proposed defense is a Sybil detection mechanism based on the novel concept of {\em
  proximity graph}.  Specifically, we leverage physical proximity between
real devices to create {\em collocation edges}, which act as secure
attestations of shared physical presence.  In a proximity graph, nodes
are Waze devices (uniquely identified by an account username and password on the
server side). They perform secure peer-to-peer location authentication
with the Waze app running in the background. An edge is established if the
proximity authentication is successful. 

Because Sybil devices are scripted software, they are highly unlikely to come
into physical proximity with real devices.  A Sybil device can only form
collocation edges with other Sybil devices (with coordination by the
attacker) or the attacker's own physical devices. The resulting graph
should have only very few (or no) edges between virtual
devices and real users (other than the attacker). 
Leveraging prior work on Sybil detection in social networks, groups of
Sybils can be characterized by the few ``attack edges'' connecting them to the
rest of the graph, making them identifiable through
community-detection algorithms~\cite{sybilroundup-sigcom10}.

We use {\em a very small number} of trusted nodes only to bootstrap trust
in the graph. We assume a small number of infrastructure access points are known to
Waze servers, {\em e.g.}, hotels and public WiFi networks associated
with physical locations stored in IP-location databases (used for
geolocation by Apple and Google). Waze also can work with
merchants that own public WiFi access points ({\em e.g.}, Starbucks). These
infrastructures are trusted nodes (we assume trusted nodes don't
collude with attackers). Any Waze device that
communicates with the Waze server under their IPs (and
reports a GPS location consistent with the IP) automatically creates a
new collocation edge to the trusted node. 

Our Sybil defense contains two key steps. First, we build a
proximity graph based on the ``encounters'' between Waze
users (\S\ref{sec:wifi}). 
Second, we detect Sybils
based on the trust propagation in proximity graph (\S\ref{sec:sybil}).

\secspace
\subsection{Peer-based Proximity Authentication}
\label{sec:wifi}
\secspace
To build the proximity graph, we first need a reliable method to 
verify the {\em physical} collocation of mobile devices.  We cannot rely
on GPS reports since attackers can forge arbitrary GPS coordinates, or
Bluetooth based device ranging~\cite{applaus_zhu2013} because the
coverage is too short ($<$10 meters) for vehicles. Instead, we
consider a challenge-based proximity authentication
method, which leverages the limited transmission range of WiFi radios.

\para{WiFi Tethering Challenge.} 
We use the smartphone's WiFi radio to implement a proximity challenge between
two Waze devices. Because WiFi radios have limited ranges ($<$250
meters for 802.11n~\cite{Tjensvold07})), two Waze devices must be in
physical proximity to complete the challenge. Specifically, we (or the Waze
server)  instruct one device to enable WiFi tethering and broadcast beacons with an SSID
provided by the Waze server, {\em i.e.}, a randomly generated,
time-varying bit string. This bit string cannot be forged \fixme{by other users or used
to re-identify a particular user}. The second device proves its proximity to
the first device by returning the SSID value heard over the air to the
Waze server. 

The key concerns of this approach are whether the WiFi link between two vehicles is
stable/strong enough to complete the challenge, and whether the
separation distance is long enough for our needs. This concern is valid
given the high moving speed, potential signal blockage from vehicles'
metal components, and the low transmit power of smartphones. We
explore these issues with detailed measurements on real mobile devices.

{\em First}, we perform measurements on stationary vehicles to study the
joint effect of blockage and limited mobile transmit power. We put two
Android phones into two cars (with windows and doors closed), one running WiFi
tethering to broadcast beacons and the other scanning for
beacons. Figure~\ref{fig:static1} plots the WiFi beacon strength at different
separation distances. We see that the above artifacts make the signal
strength drop to -100 dBm before the distance reaches 250 meters.
In the same figure, we also plot the probability of successful beacon
decoding (thus challenge completion) across 400 attempts within 2 minutes. It
remains 100\% when the two cars are separated by $<$80 meters, and drops
to zero at 160 meters.

{\em Next}, we perform driving experiments on a highway at normal traffic hours in
the presence of other vehicles. The vehicles travel at speeds
averaging 65 mph. During driving, we are able to vary the distance
between the two cars, and use recorded GPS logs to calculate the
separation distance.  
Figure~\ref{fig:drive2} shows that while WiFi
signal strength fluctuates during our experiments, the probability of beacon
decoding remains very high at 98\% when the separation is less than 80 meters
but drops to $<$10\% once the two cars are more than 140 meters apart.

Overall, the results suggest the proposed WiFi tethering
challenge is a reliable method for proximity authentication for our
system. In practice, Waze can start the challenge when detecting the two vehicles are
within the effective range, {\em e.g.}, 80 meters. Since the WiFi
channel scan is fast, {\em e.g.}, 1-2 seconds to do a full channel
scan in our experiments, this challenge can be
accomplished quickly with minimum energy cost on mobile
devices. It is easy to implement this scheme using existing
APIs to control WiFi radio to open tethering ({\tt setWifiApEnabled}
API in Android). 

\para{Constructing Proximity Graphs.}  In a proximity graph, each
node is a Waze device, and an edge indicates the two users come into physical
proximity, {\em e.g.}, 80 meters, within a predefined time window. 
The resulting graph is undirected but weighted based on the number
of times the two users have encountered.
Using weighted graph makes it harder for Sybils to blend into the
normal user region. Intuitively, real users will get more weights on their
edges as they use Waze over time. For attackers, in order to blend in
the graph, they need to build more weighted attack edges to real users
(higher costs).

\fixme{ This approach should not introduce much energy consumption to users'
  phones. First, Waze server does not need to trigger collocation
  authentication every time two users are in close proximity. Instead, the
  proximity graph will be built up over time. A user only need to
  authenticate with other users occasionally, since we can require that
  device authentication expires after a moderate time period ({\em e.g.,}
  months) to reduce the net impact on wireless performance and energy
  usage. Second, since the process is triggered by the Waze server, Waze can
  can use WiFi sensing from devices to find ``opportunistic'' authentication
  times that minimize impact on performance and energy. Waze can also use one
  tether to simultaneously authenticate multiple colocated devices within an
  area. This further reduces authentication overhead, and avoids performance
  issues like wireless interference in areas with high user density.  }

\secspace
\subsection{Graph-based Sybil Detection}
\label{sec:sybil}
\secspace
We apply graph-based Sybil detection algorithms to detect Sybils in Waze
proximity graph. Graph-based Sybil detectors~\cite{sybilguard, sybillimit,
  sybilroundup-sigcom10, sybilinfer, sybilrank} were originally
proposed in social networks. They all rely on the key assumption that Sybils have
difficulty to form edges with real users, which results in a
sparse cut between the Sybil and non-Sybil regions in the social
graph. Because of the limited number of ``attack edges''
between Sybils and non-Sybils, a random walk from non-Sybil region
has a higher landing probability to land on a non-Sybil node than a Sybil
node. Our proximity graph holds the same assumption that these
algorithms require---with the WiFi proximity authentication, it's
difficult for Sybil devices (ghost riders) to build attack edges to real
Waze users.

\para{SybilRank.}  Among available algorithms, we use
SybilRank~\cite{sybilrank}. Compared to its counterparts
(SybilGuard~\cite{sybilguard}, SybilLimit~\cite{sybillimit} and
SybilInfer~\cite{sybilinfer}), SybilRank achieves higher accuracy at a lower
computational cost. At the high-level, its counterparts 
% (SybilGuard, SybilLimit and SybilInfer) 
need to perform {\em actual random walks}, which is very costly and
yet often gives incomplete views of the graph. Instead, SybilRank uses {\em power
  iteration}~\cite{poweriteration2004} to compute the random walk landing
probability for all nodes. This significantly boosts the algorithm
accuracy and speed. Furthermore, SybilRank has a better tolerance on
community structures in the non-Sybil region (for using multiple trusted
nodes), making it more suitable for real-world graphs.

As context, we briefly describe
how SybilRank works and refer readers to~\cite{sybilrank} for more
details. SybilRank ranks the nodes based on how likely
they are Sybils. The algorithm starts with multiple trusted
nodes in the graph. It iteratively computes the landing probability for short random walks
(originated from trusted nodes) to land on all other nodes. The landing
probability is normalized by the node's degree, which acts as the
trust score for ranking. Intuitively, short random walks from trusted
nodes are very unlikely to traverse the few attack edges to
reach Sybil nodes, thus the ranking scores of Sybils should be
lower. For Sybil detection, Waze can set a cutoff threshold
on the trust score, and label the tail of the ranked list as Sybils. 

The original SybilRank works on unweighted social graphs. We
modified it to work on our weighted proximity graph: when a node
propagates trust (or performs random walks) to its neighbors, instead
of splitting the trust equally, it distributes proportionally based on
the edge weights. This actually makes it harder for Sybils to evade
SybilRank---they will need to build more high-weight attack edges to
real users to receive trust.

\secspace
\section{Countermeasure Evaluation}
\label{sec:eval}
\secspace
We use simulations to evaluate the effectiveness of our proposed
defense. We focus on evaluating the feasibility and cost for attackers to
maintain a large number of Sybils after the Sybil detection is in place. We
quantify the cost by the number of attack edges 
% (in the proximity graph)
a Sybil must establish with real users. In practice, this translates into the
effort taken to physically drive around and use physical devices (with
WiFi radios) per Sybil to complete proximity authentication. 
In the following, we first describe our simulation setup, and then
present the key findings and their implications on Waze. 

\begin{figure*}[t]
\centering
\begin{minipage}{0.65\textwidth}
\begin{center}
\mbox{
\subfigure[Sybil inner connection avg. degree = 5]{
    \includegraphics[width=0.47\textwidth]{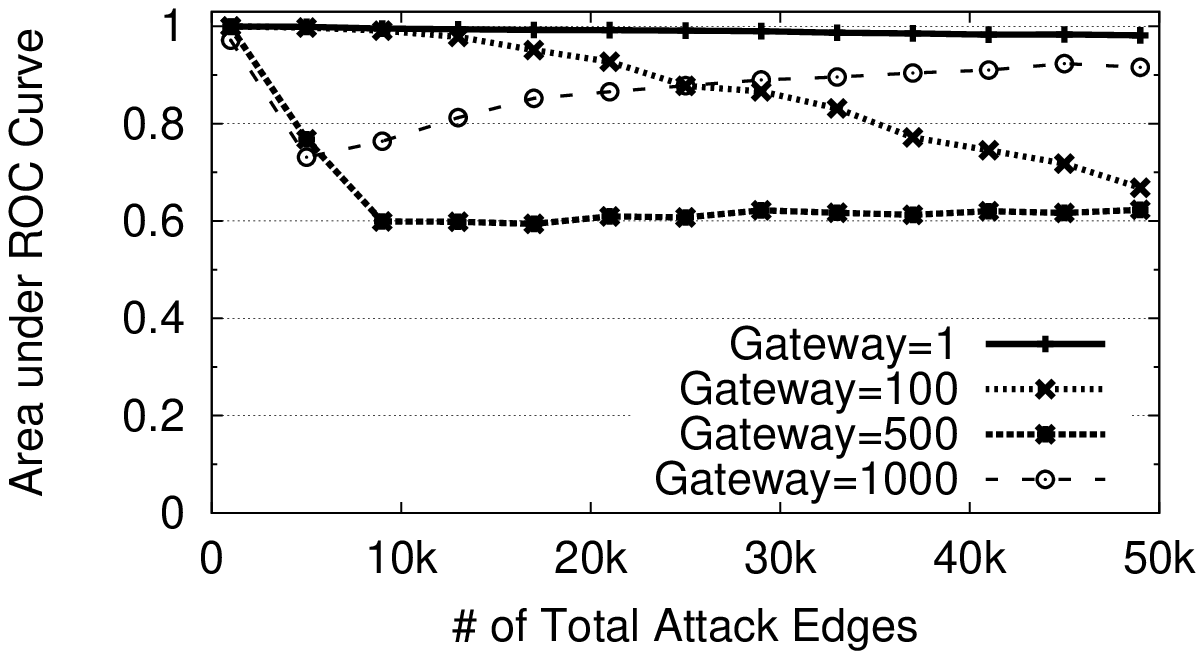}
	\label{fig:sybil_inner1}
 }
\hspace{0.1in}
 \subfigure[Sybil inner connection avg. degree = 10]{
	\includegraphics[width=0.47\textwidth]{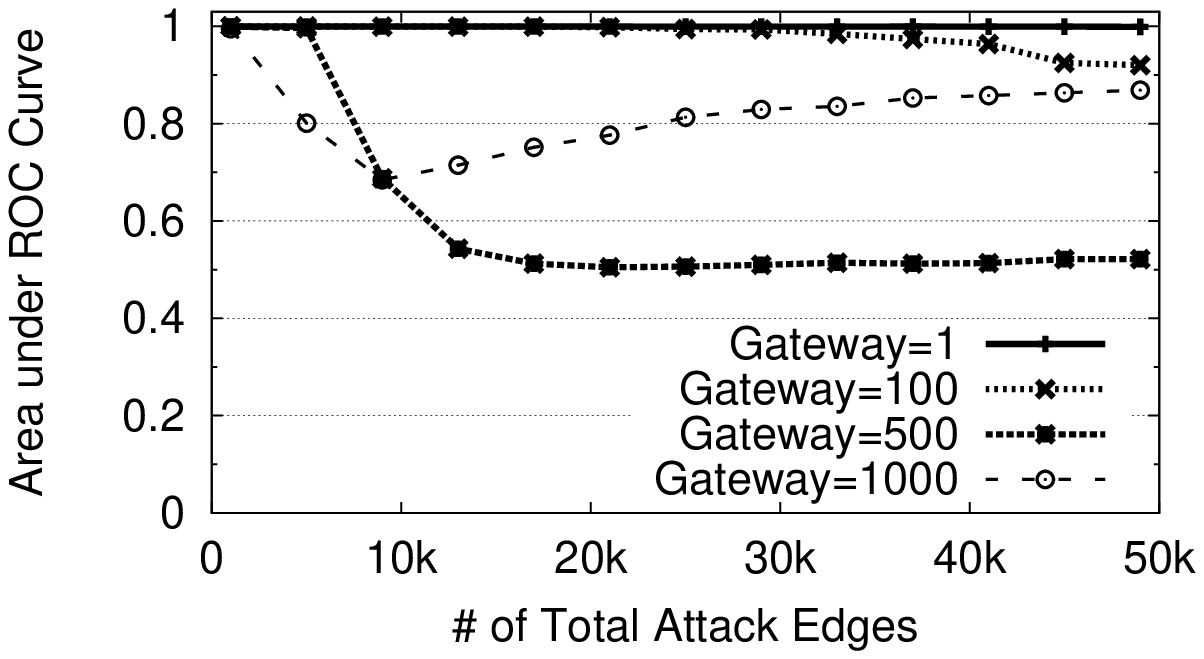}
	\label{fig:sybil_inner2}
 }
 }
\vspace{-0.05in}
\caption{AUC with respect to number of attack
  edges, where Sybils form power-law inner connections.}
\label{fig:sybil_inner}
\vspace{-0.05in}
\end{center}
\end{minipage}
\vspace{-0.05in}
\hfill
\begin{minipage}{0.32\textwidth}
 \centering
	\includegraphics[width=0.95\textwidth]{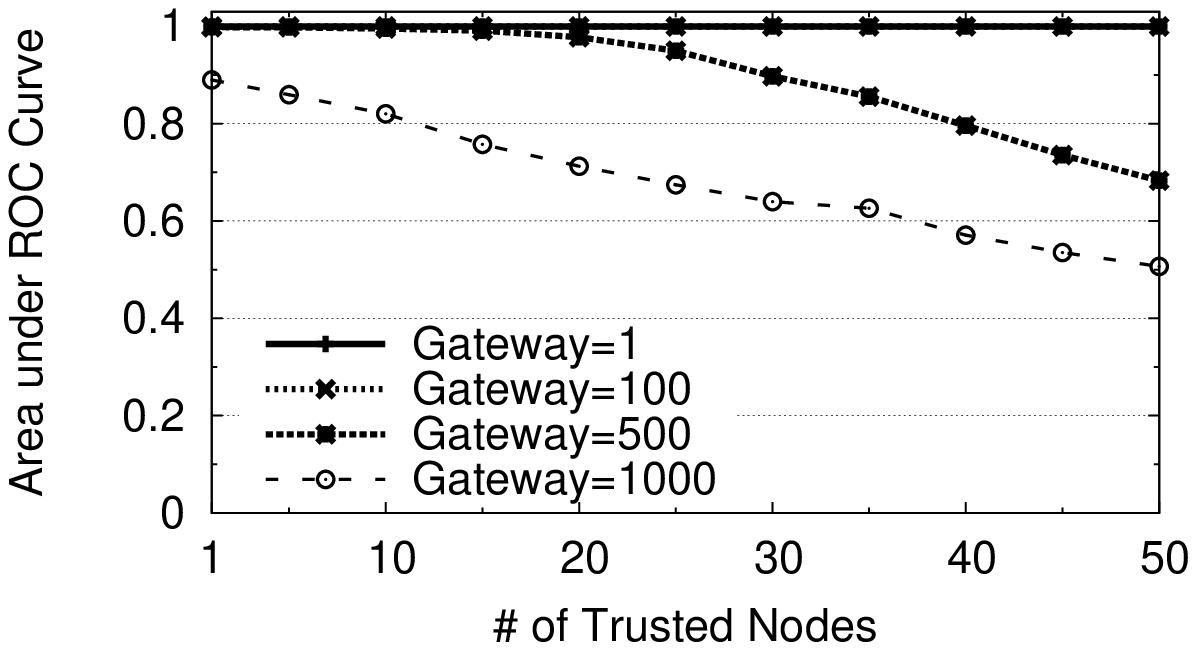}
        \vspace{-0.05in}
        \caption{Impact of \# of trusted nodes (average degree
          =10 for Sybil region; 5K attack edges).}
	\label{fig:sybil_seed}
\end{minipage}
\vspace{-0.05in}
\end{figure*}

\secspace
\subsection{Evaluation Setup}
\secspace
We first discuss how we construct a synthetic proximity graph for our
evaluation, followed by the counter strategies taken by attackers to
evade detection. Finally, we describe the evaluation
metrics for Sybil detection. 

\para{Simulating Proximity Graphs.} We use well-known models on human
encountering to create synthetic proximity graphs.  This is because, to the
best of our knowledge, there is no public {\em per-user} mobility dataset with sufficient
scale and temporal coverage to support our evaluation. Also, directly
crawling large-scale, per-user mobility trace from Waze can lead to
questionable privacy implications, and thus we exclude this option.

Existing literatures~\cite{socialmobile07, vehicular14,
  wowmom09, DTN:2010, Liu:2012} all suggest that human (and vehicle) encounter
patterns display strong scale-free and ``small-world''
properties~\cite{barabasi-1999}. Thus we follow the methodology of~\cite{socialmobile07} to
simulate a power-law based encounter process among Waze users. Given a user
population $N$, we first assign each user an encounter probability
following a power-law distribution ($\alpha=$2 based on the empirical
values \cite{socialmobile07, clauset2009power}). We then simulate
user encounter over time, by adding edges to the graph based on the
joint probability of the two nodes.

For our evaluation, we produce a proximity graph for $N=10000$ normal users and use the
snapshot when 99.9\% of nodes are connected. Note that as the graph
gets denser over time, it is harder for Sybils to blend into
normal user regions. We use this graph to simulate the lower-bound performance of
Sybil detection.\footnote{\fixme{Validated by experiments: a denser, 99.99\%
  connected graph can uniformly improve Sybil detection accuracy.}} 
\fixme{As a potential limitation, the simulated graph parameters might
  be different for different cities of Waze. Thus we don’t
  claim our reported numbers will exactly match what Waze
  produces. The idea is that Waze can follow our methodology to run
  the same experiments on their real graphs.}

\para{Attacker Models.} In the presence of Sybil detection, an attacker will
try mixing their Sybils into the proximity graph.
% without being detected
We consider the following strategies: 
\begin{packed_enumerate}
\item {\bf Single-Gateway} -- An attacker first takes one
  Sybil account (as the gateway) to build attack edges to normal
  users. Then the attacker connects the remaining Sybils to this
  gateway. In practice, this means the attacker only needs to
  take one physical phone to go out and encounter normal users.
% to drive around

\item {\bf Multi-Gateways} -- An attacker distributes
  the attack edges to multiple gateways, and then evenly spreads the other
  Sybils across the gateways. This helps the Sybils to blend in with normal
  users. The attacker pays an extra cost in terms of using multiple real
  devices to build attack edges. 
\end{packed_enumerate}
The attacker also builds edges among its own Sybils. This incurs no
additional cost since Sybils can easily collude to pass proximity
authentication, but introduces key benefits. {\em First}, it makes
Sybils' degree distribution appear more legitimate. {\em Second}, it can
potentially increase Sybils' trust score: when a random walk
reaches one Sybil node, its edges to the fellow Sybils help to
sustain the random walk within the Sybil region.
 % and boost Sybils' trust score. 
In our simulation, we follow the scale-free distribution to add edges
among Sybils mimicking normal user region (we did not use a fully
connected network between Sybils since it is more easily detectable). 

\para{Evaluation Metrics.} To evaluate Sybil detection efficacy, we use
the standard false positive (negative) rate, and the Area under the
Receiver Operating Characteristic curve (AUC) used
by SybilRank~\cite{sybilrank}. AUC represents the probability that
SybilRank ranks a random Sybil node lower than a random non-Sybil node. Its value
ranges from 0 to 1, where 1 means the ranking is perfect (all Sybils are ranked
lower than non-Sybils),  0 means the ranking is always flipped,
%  (all Sybils are ranked higher than non-Sybils)
and 0.5 matches the result of random guessing. Compared to false
positive (negative) rates, AUC is independent of the cutoff threshold,
and thus comparable across experiment settings.

\begin{figure*}[t]
\centering
\begin{minipage}{0.65\textwidth}
\begin{center}
\mbox{
\subfigure[False Positive Rate]{
    \includegraphics[width=0.47\textwidth]{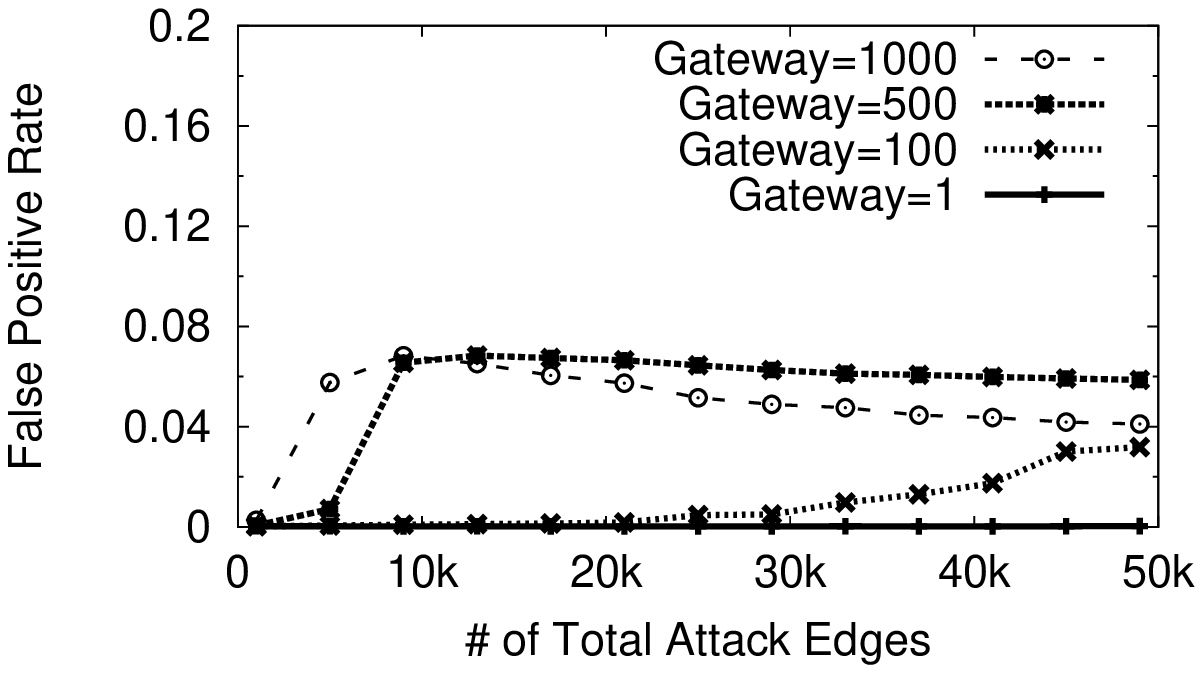}
	\label{fig:sybil_fp}
 }
\hspace{0.1in}
 \subfigure[False Negative Rate]{
	\includegraphics[width=0.47\textwidth]{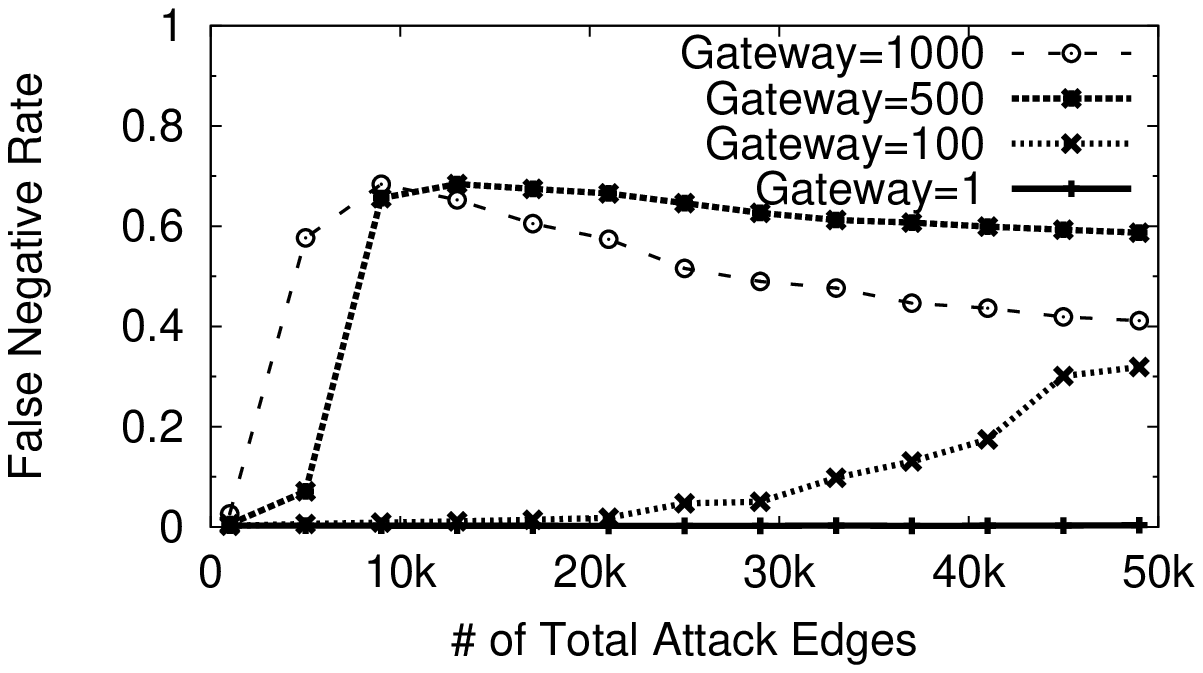}
	\label{fig:sybil_fn}
 }
 }
\vspace{-0.05in}
\caption{Detection error rates with respect to number of attack
  edges. We set average degree =10 for Sybils' power-law inner connections.}
\label{fig:sybil_error}
\vspace{-0.05in}
\end{center}
\end{minipage}
\vspace{-0.05in}
\hfill
\begin{minipage}{0.32\textwidth}
 \centering
	\includegraphics[width=1\textwidth]{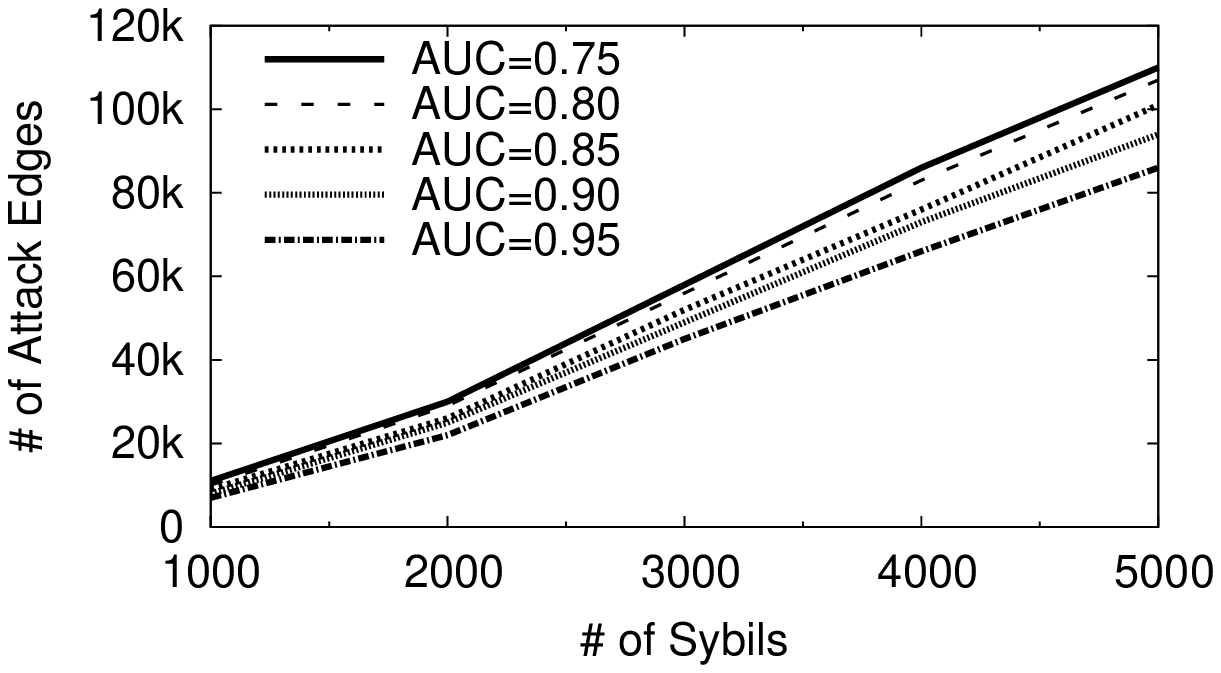}
        \vspace{-0.15in}
        \caption{\# of attack edges needed to maintain $x$ Sybil
          devices with respect to different AUC level.}
	\label{fig:cost}
\end{minipage}
\vspace{-0.05in}
\end{figure*}

\subsection{Results}
\para{Accuracy of Sybil Detection.}  We assume the attacker seeks to embed
1000 Sybils into the proximity graph. We use either single- or 
multi-gateway approaches to build attack edges on the proximity graph by
connecting Sybils to randomly chosen normal users. We then add edges between
Sybil nodes, following the power-law distribution and producing an average
weighted degree of either 5 or 10 (to emulate different Sybil subgraph
density). We randomly select 10 trusted nodes to bootstrap trust for
SybilRank and run it on the proximity graph. We
repeat each experiment 50 times. 

Figure~\ref{fig:sybil_inner} shows that the Sybil
detection mechanism is highly effective.  For attackers of the single-gateway
model, the AUC is very close to 1 ($>0.983$), indicating Waze can
identify almost all Sybils even after the attacker established a
large number of attack edges, {\em e.g.}, 50000. Meanwhile, the multi-gateway
method helps attackers add ``undetected'' Sybils, but the
number of gateways required is significant. For example, to
maintain 1000 Sybils, {\em i.e.}, by bringing down AUC to 0.5, the
attacker needs at least 500 as gateways. In practice, this means
wardriving with 500+ physical devices to meet real users, which is a
significant overhead. 

Interestingly, the 1000-gateway result (where every
Sybil is a gateway) shows that, at certain point, adding more attack edges
can actually hurt Sybils. 
This is potentially due to the fact that SybilRank uses node degree to
normalize trust score. For gateways that connect to both normal users
and other Sybils, the additional ``trust'' received
by adding more attack edges cannot compensate the penalty of degree
normalization.

For a better look at the {\em detection accuracy},  we convert the AUC
in Figure~\ref{fig:sybil_inner2} to false positives (classifying
real users as Sybils) and false negatives (classifying Sybils as real
users). For simplicity, we set a cutoff value to mark the bottom 10\% of the
ranked nodes as Sybils.\footnote{This cutoff value is only to convert the
error rate. In practice, Waze can optimize this value based on the
trust score or manual examination.} 
As shown in
Figure~\ref{fig:sybil_error}, SybilRank is highly accurate to detect
Sybils when the number of gateways is less than 100. Again 100
gateways incurs high cost in practice.

Next we quickly examine the impact of trusted nodes to Sybil
detection. Figure~\ref{fig:sybil_seed} shows a small
number of trusted node is enough to run SybilRank. Interestingly,
adding more trusted nodes can slightly hurt Sybil detection, possibly
because it gives the attacker (gateways) a higher chance to receive
trust. In practice, multiple trusted nodes can help SybilRank overcome
potential community structures in proximity graph ({\em e.g.}, users
of the same city form a cluster). So Waze should place
trusted nodes accordingly to cover geographic clusters.

\para{Cost of Sybil Attacks.} Next, we infer the rough cost of attackers
on implementing successful Sybil attacks. For this we look at the number of
attack edges required to successfully embed a given number of Sybils.  Our
experiment assumes the attacker uses 500 gateways and builds power-law
distributed inner connections with average
degree$=$10. Figure~\ref{fig:cost} shows the number of attack edges required
to achieve a specific AUC under SybilRank as a function of the target number
of Sybils. We see that the attack edge count increases linearly with the
Sybil count. The cost of Sybil attack is high: to maintain 3000 Sybils, the
attacker must make 60,000 attack edges to keep AUC below 0.75, and spread
these attack edges across 500 high-cost gateways. 

\fixme{
\para{Smaller Sybil Groups.} Finally, we examine how effective
our system is in detecting much smaller Sybil groups. We test Sybil groups with
size of 20, 50 and 100 using a single-gateway approach. We configure
50K attacking edges for Sybils with inner degree = 10. The resulting
AUC of Sybil detection is 0.90, 0.95 and 0.99 respectively. % Not too
% surprisingly, smaller Sybil groups are more challenging to detect. 
This confirms our system can effectively identify small
Sybil groups as well.}

\subsection{Implications on Waze}
\label{sec:imp}

These results show that our Sybil detection method is highly effective. It
significantly increases the cost (in purchasing physical devices and time
spent actually driving on the road) to launch Sybil attacks. Also, SybilRank
is scalable enough for large systems like Waze.  A social network with tens
of millions of users has been running SybilRank on Hadoop
servers~\cite{sybilrank}.

In addition to Sybil detection, Waze can incorporate other mechanisms
to protect its users. We briefly describe a few key ideas, but
leave the integration with our approach to future work.
{\em First}, IP verification: when a user claims she is driving, Waze
 can examine whether her IP is a mobile IP that belongs to a
 valid cellular carrier or a suspicious web proxy. However,
 this approach is ineffective if dedicated attackers route the attack
 traffic through a cellular data plan. {\em Second}, strict rate limit: with that, attackers will
need to run more Sybil devices to implement the same attack. 
{\em Third}, verifications on account registration: this needs to be
handled carefully since email/SMS based verification
can be bypassed using disposable email or phone
numbers~\cite{phone_ccs14}.
\fixme{
{\em  Finally}, detecting extremely inconsistent GPS/event
reports. The challenge, however, is to distinguish
honest reports from the fake ones since attacker can easily outvote real
users. If Waze chooses to ignore all the inconsistent reports, it will lead to
DOS attack where attackers disable the service with inconsistent data.}

\section{Broader Implications}
\secspace
\label{sec:broad}
While our experiments and defenses have focused strictly on Waze,
our results are applicable to a wider range of mobile
applications that rely on geolocation for user-contributed content and
metadata.  Examples include location based check-in and review services
(Foursquare, Yelp), crowdsourced navigation systems (Waze, Moovit),
crowdsourced taxi services (Uber, Lyft), mobile dating apps (Tinder, Bumble)
and anonymous mobile communities (Yik Yak, Whisper).

These systems face two common challenges exposing them to potential
attacks. First, our efforts show that it is difficult for app developers
to build a truly secure channel between the app and the server.  There are
numerous avenues for an attacker to reverse-engineer and mimic an app's API
calls, thereby creating ``cheap'' virtual devices and launching Sybil
attack~\cite{Sybil}. Second, there are no deployed mechanisms to
authenticate location data ({\em e.g.}, GPS report). Without a secure
channel to the server and authenticated location, these mobile apps
are vulnerable to automated attacks ranging from nuisance (prank
calls to Uber) to malicious content attacks (large-scale rating
manipulation on Yelp).

To validate our point, we run a quick empirical analysis on a broad
class of mobile apps to understand how easy it is to reverse-engineer their APIs and
inject falsified data into the system. We pick one app from
each category including Foursquare, Uber, Tinder and Yik Yak (an incomplete
list). We find that, although all the listed apps use TLS/SSL to encrypt
their network traffic, their APIs can be fully
exposed by the method in \S\ref{sec:scale}. For each app, we were able to
build a light-weight client using python script, and feed
arbitrary GPS to their key function calls. For example, with forged
GPS, a group of Foursquare clients can deliver large volumes of
check-ins to a given venue without physically visiting it; On Uber, one can
distribute many virtual devices as sensors, and passively
monitor and track all drivers (and their passengers) within a large area (see
\S\ref{sec:track}). Similarly for Yik Yak and Tinder, the virtual devices
make it possible to perform wardriving in a given location area to post and
collect anonymous Yik Yak messages or Tinder profiles. \fixme{In
  addition, apps like Tinder also display the geographical distance to a nearby
  user ({\em e.g.}, 1 mile). Attacker can use multiple virtual devices
  to measure the distance to the target user, and ``triangulate'' that
  user's exact location~\cite{whisperIMC14}.} There are possible app-specific
defenses, and we leave their design and evaluation to future work.

\section{Disclosure and Impact} 
Before the first writeup of our work, we sought to inform the Google Waze
team of our results. We first used multiple existing Google contacts on the
security and Android teams. When that failed to reach the Waze team, we got
in touch with Niels Provos, who then relayed information about our project to
the Waze team.

Through our periodic tests of the Waze app, we noticed recent updates made
significant changes to how the app reports user location data to the server
(and other users). In the new Waze update (v4.4.0, released in April 2016),
the app only reports user GPS values when the user is actively driving
(moving at a moderate/fast rate of speed). GPS tracking stops when a user is
walking or standing still. In addition, Waze automatically shuts down if the
user puts it in the background, and has not driven for a while. To resume
user tracking (GPS reporting), users must manually bring the app to the
foreground. Finally, Waze now hide users' starting and destination locations
of their trips. 

While online documentation claims that these optimizations are to reduce
energy usage for the Waze app, we are gratified by the dramatic steps taken
to limit user tracking and improve user privacy. These changes
dramatically reduce the amount of GPS data sent to the server (and made
available to potential attackers through the API). By our estimates, the
update reduces the amount of GPS tracking data for a typical user by nearly a
factor of 10x. In addition, removing the first and last GPS values of a trip
means that it is significantly harder to track a user through multiple
trips. Previously, users could be tracked across new Waze sessions, despite
new per-session identifiers, by matching the destination point of one trip
with the starting point of the next.

We note that while Waze has taken significantly steps to improve user
privacy, users can still be tracked while they are actively using the
app. More importantly, the attack we identified here can still wreak havoc
with a wide range of mobile apps, and Sybil devices are a real challenge
still in need of practical solutions. We hope our work spurs future work to
address this problem. 

\section{Related Work}
\label{sec:related}

\para{Security in Location-based Services.} 
Location-based services face various threats, ranging from rogue users 
reporting fake GPS~\cite{mass12foursquare,ICDCS11foursqaure}, to 
malicious parties compromising user privacy~\cite{location13,krumm2007inference,krumm2009survey}.
A related study on Waze~\cite{sinai14} demonstrated that small-scale 
attacks can create traffic jams or track user icons, with up to 15 
mobile emulators. Our work differs in two key aspects. First, we show 
that it's possible to reverse engineer its APIs, enabling  
light-weight Sybil devices (simple scripts) to replace full-stack  
emulators. This increase the scale of potential attacks by orders of 
magnitude, to thousands of Waze clients per commodity 
laptop. The impact of thousands of virtual vehicles is qualitatively  
different from 10-15 mobile simulators. \fixme{Second}, as possible 
defenses, \cite{sinai14} cites known tools such as phone number/IP verification, 
or location authentication with cellular towers, which have limited 
applicability (see \S\ref{sec:defense}).  
In contrast, we propose a novel proximity graph approach 
to detect and constrain the impact of virtual devices. 

Researchers have proposed to preserve user location privacy against 
map services such as Waze and Google. \fixme{Earlier studies apply
  location cloaking by adding noise to the GPS reports~\cite{mobisys03}. Recent
work use zero-knowledge~\cite{blackhat13waze} and differential 
privacy~\cite{haze13} to preserve the location privacy of individual 
users, while maintaining user accountability and the accuracy of
aggregated statistics. } Our work differs by focusing on the attacks against the
map services. 

\para{Mobile Location Authentication.} Defending against 
forged GPS is challenging. One direction is to authenticate user
locations using wireless infrastructures: WiFi
APs~\cite{hotmobile10locpro1,hotmobile09locproof}, 
cellular base stations~\cite{hotmobile10locpro1,hotmobile09locproof} 
and femtocells~\cite{femtocells_brassil2014}.  
Devices must come into physical proximity to these infrastructures to
be authenticated. But it requires cooperation among a wide range of
infrastructures (also modifications to their software/hardware), which is impractical for 
large-scale services like Waze. Our work  
only uses a small number of trusted infrastructures to bootstrap, and
relies on peer-based trust propagation to achieve coverage. 
Other researchers have proposed ``peer-based'' methods to 
authenticate collocated mobile devices~\cite{link_talasila10,stamp_icnp13,applaus_zhu2013,smile2009, 
  wifiproxNDSS11}. Different from existing work, we use peer-based 
collocation authentication to build proximity graphs for Sybil 
detection, instead of directly authenticating a device's physical
location.  

\para{Sybil Detection. }  
Sybil detection has been studied in P2P networks~\cite{Sybil} and
online  social networks~\cite{sybilroundup-sigcom10, sybils-ndss13,
    clickstream13}. The most popular approach is graph-based where the 
  key assumption is that Sybils have difficulty to connect to real 
  users~\cite{sybilrank,sybilinfer, sumup,sybillimit, sybilguard}.  
Thus Sybils would form a well-connected subgraph that has a small
quotient-cut from the non-Sybil region. Our work  
constructs a proximity graph that holds the same assumption, and 
applies Sybil detection algorithm to locate ghost riders in Waze.  
We differ from~\cite{sybilrank} in the graph used and the attack 
models.

\section{Conclusion}
We describe our efforts to identify and study a range of
attacks on crowdsourced map services.  We identify a range of single and
multi-user attacks, and describe techniques to build and control groups of
virtual vehicles (ghost riders) to amplify these attacks.  Our work shows
that today's mapping services are highly vulnerable to software agents
controlled by malicious users, and both the stability of these services and
the privacy of millions of users are at stake.
While our study and experiments focus on the Waze system, we believe the
large majority of our results can be generalized to crowdsourced apps as a
group.  We propose and validate a suite of techniques that help 
services build proximity graphs and use them to effectively detect Sybil
devices.

Throughout this work, we have taken active steps to isolate our experiments
and prevent any negative consequence on real Waze users. We also used
our existing Google/Waze contacts to inform Waze team of our
results. More details on IRB, ethics and disclosure are contained in Appendix A.

\section*{Acknowledgments}
We would like to thank our shepherd Z. Morley Mao and the anonymous reviewers
for their comments. This project was supported by NSF grants
CNS-1527939 and CNS-1224100. Any opinions, findings, and
conclusions or recommendations expressed in this material are those of the
authors and do not necessarily reflect the views of any funding agencies.

\section*{Appendix A--- IRB, Ethics}
\label{sec:app_a}
Our study was reviewed and approved by our local IRB. Prior to doing any
real measurements on the system, we submitted a human
subject protocol for approval by our institutional IRB. The protocol
was fully reviewed for ethics and privacy risks, and the response was
our study can be exempt. We put the request into our IRB system and
began our work. Then the confirmation of exemption arrived and our
study received the IRB approval under protocol~\# COMS-ZH-YA-010-7N.
 
As described in the paper, we are very aware of the potential impact on real
Waze users from any experiments.  We took very careful precautions to ensure
that our experiments will not negatively impact Waze servers or Waze
users. In particular, we conducted numerous measurements of diverse traffic
regions (read-only) to locate areas of extremely low traffic density.  We
chose experiment locations where user population density is extremely low
(unoccupied roads), and only perform experiments at low-traffic hours, {\em
  e.g.} between 3am and 5am. During experiments, we continuously scan the
entire region including our experimental area and neighboring regions, to
ensure no other Waze users (except our own accounts) are within miles of the
test area.  If any Waze users are detected, we immediately terminate any
running experiments.  We took care to limit congestion tests to areas with
lots of local route redundancy, thus we would not affect the routing of any
long distance trips (e.g. taking highway 80 because the 101 was congested).
Finally, while we cannot detect invisible users in our test area, we have
taken every precaution to only test on roads and times that show very little
traffic, e.g. low population areas at 4am local time.  We believe in practice, invisible
users make up a small subset of the Waze population, because they cannot send
reports or message other users (effectively removing most/all of the social
functionality in Waze), and Waze resets the invisible setting every time the
app is opened~\cite{waze:config}.

\balance
\bibliographystyle{abbrv}
\bibliography{zhao,astro}

\begin{thebibliography}{10}

\bibitem{waze:config}
{About Waze: Privacy}.
\newblock \url{https://support.google.com/waze/answer/6071193?hl=en}.

\bibitem{charles}
{Charles Proxy}.
\newblock \url{http://www.charlesproxy.com}.

\bibitem{genymotion}
{GenyMotion Emulator}.
\newblock \url{http://www.genymotion.com}.

\bibitem{monkeyrunner}
{Monkeyrunner}.
\newblock \url{http://developer.android.com/tools/help/
  monkeyrunner_concepts.html}.

\bibitem{osrm}
{Open Source Routing Machine (OSRM)}.
\newblock \url{http://map.project-osrm.org}.

\bibitem{barabasi-1999}
A.-L. Barabasi and R.~Albert.
\newblock Emergence of scaling in random networks.
\newblock {\em Science}, 286, 1999.

\bibitem{femtocells_brassil2014}
J.~Brassil, P.~K. Manadhata, and R.~Netravali.
\newblock Traffic signature-based mobile device location authentication.
\newblock {\em IEEE Transactions on Mobile Computing}, 13(9):2156--2169, 2014.

\bibitem{haze13}
J.~W.~S. Brown, O.~Ohrimenko, and R.~Tamassia.
\newblock Haze: Privacy-preserving real-time traffic statistics.
\newblock In {\em Proc. of SIGSPATIAL}, 2013.

\bibitem{sslsp:2014}
C.~Brubaker, S.~Jana, B.~Ray, S.~Khurshid, and V.~Shmatikov.
\newblock Using frankencerts for automated adversarial testing of certificate
  validation in ssl/tls implementations.
\newblock In {\em Proc. of IEEE S\&P}, 2014.

\bibitem{sybilrank}
Q.~Cao, M.~Sirivianos, X.~Yang, and T.~Pregueiro.
\newblock Aiding the detection of fake accounts in large scale social online
  services.
\newblock In {\em Proc. of {NSDI}}, 2012.

\bibitem{mass12foursquare}
B.~Carbunar and R.~Potharaju.
\newblock You unlocked the mt. everest badge on foursquare! countering location
  fraud in geosocial networks.
\newblock In {\em Proc. of MASS}, 2012.

\bibitem{clauset2009power}
A.~Clauset, C.~R. Shalizi, and M.~E. Newman.
\newblock Power-law distributions in empirical data.
\newblock {\em SIAM review}, 51(4):661--703, 2009.

\bibitem{vehicular14}
F.~Cunha, A.~C. Viana, R.~A.~F. Mini, and A.~A.~F. Loureiro.
\newblock Is it possible to find social properties in vehicular networks?
\newblock In {\em Proc. of ISCC}, 2014.

\bibitem{sybilinfer}
G.~Danezis and P.~Mittal.
\newblock Sybilinfer: Detecting sybil nodes using social networks.
\newblock In {\em Proc of {NDSS}}, 2009.

\bibitem{location13}
Y.-A. de~Montjoye, M.~Verleysen, and V.~D. Blondel.
\newblock Unique in the crowd: The privacy bounds of human mobility.
\newblock {\em Scientific Reports}, 3, 2013.

\bibitem{Sybil}
J.~R. Douceur.
\newblock The {Sybil} attack.
\newblock In {\em Proc. of {IPTPS}}, 2002.

\bibitem{Decomple2011}
W.~Enck, D.~Octeau, P.~McDaniel, and S.~Chaudhuri.
\newblock A study of android application security.
\newblock In {\em Proc. of USENIX Security}, 2011.

\bibitem{SSLpinning2013}
S.~Fahl, M.~Harbach, H.~Perl, M.~Koetter, and M.~Smith.
\newblock Rethinking ssl development in an appified world.
\newblock In {\em Proc. of CCS}, 2013.

\bibitem{waze-google3}
V.~Goel.
\newblock Maps that live and breathe with data.
\newblock The New York Times, June 2013.

\bibitem{waze-google1}
Google.
\newblock Google maps and waze, outsmarting traffic together.
\newblock Google Official Blog, June 2013.

\bibitem{mobisys03}
M.~Gruteser and D.~Grunwald.
\newblock Anonymous usage of location-based services through spatial and
  temporal cloaking.
\newblock In {\em Proc. of MobiSys}, 2003.

\bibitem{ICDCS11foursqaure}
W.~He, X.~Liu, and M.~Ren.
\newblock Location cheating: A security challenge to location-based social
  network services.
\newblock In {\em Proc. of ICDCS}, 2011.

\bibitem{DTN:2010}
T.~Hossmann, T.~Spyropoulos, and F.~Legendre.
\newblock Know thy neighbor: Towards optimal mapping of contacts to social
  graphs for dtn routing.
\newblock In {\em Proc. of INFOCOM}, 2010.

\bibitem{blackhat13waze}
T.~Jeske.
\newblock Floating car data from smartphones: What google and waze know about
  you and how hackers can control traffic.
\newblock {\em Black Hat}, 2013.

\bibitem{Binomial:1988}
V.~Kachitvichyanukul and B.~W. Schmeiser.
\newblock Binomial random variate generation.
\newblock {\em Commun. ACM}, 31(2):216--222, 1988.

\bibitem{krumm2007inference}
J.~Krumm.
\newblock Inference attacks on location tracks.
\newblock In {\em Pervasive Computing}. 2007.

\bibitem{krumm2009survey}
J.~Krumm.
\newblock A survey of computational location privacy.
\newblock {\em Personal and Ubiquitous Computing}, 2009.

\bibitem{poweriteration2004}
A.~Langville and C.~Meyer.
\newblock {Deeper inside pagerank}.
\newblock {\em Internet Mathematics}, 1(3):335--380, 2004.

\bibitem{Liu:2012}
X.~Liu, Z.~Li, W.~Li, S.~Lu, X.~Wang, and D.~Chen.
\newblock Exploring social properties in vehicular ad hoc networks.
\newblock In {\em Proc. of Internetware}, 2012.

\bibitem{hotmobile10locpro1}
W.~Luo and U.~Hengartner.
\newblock Proving your location without giving up your privacy.
\newblock In {\em Proc. of HotMobile}, 2010.

\bibitem{smile2009}
J.~Manweiler, R.~Scudellari, and L.~P. Cox.
\newblock Smile: Encounter-based trust for mobile social services.
\newblock In {\em Proc. of CCS}, 2009.

\bibitem{ndss14pay}
C.~Marforio, N.~Karapanos, C.~Soriente, and K.~Capkun.
\newblock Smartphones as practical and secure location verification tokens for
  payments.
\newblock In {\em Proc. of NDSS}, 2014.

\bibitem{socialmobile07}
A.~G. Miklas, K.~K. Gollu, K.~K.~W. Chan, S.~Saroiu, K.~P. Gummadi, and
  E.~de~Lara.
\newblock Exploiting social interactions in mobile systems.
\newblock In {\em Proc. of Ubicomp}, 2007.

\bibitem{wifiproxNDSS11}
A.~Narayanan, N.~Thiagarajan, M.~Lakhani, M.~Hamburg, and D.~Boneh.
\newblock {Location Privacy via Private Proximity Testing}.
\newblock In {\em Proc. of NDSS}, 2011.

\bibitem{blackhat12ssl}
J.~Osborne and A.~Diquet.
\newblock When security gets in the way: Pentesting mobile apps that use
  certificate pinning.
\newblock {\em Black Hat}, 2012.

\bibitem{googlemap4}
B.~Reed.
\newblock Google maps becomes google’s second 1 billion-download hit.
\newblock Yahoo! News, June 2014.

\bibitem{hotmobile09locproof}
S.~Saroiu and A.~Wolman.
\newblock Enabling new mobile applications with location proofs.
\newblock In {\em Proc. of HotMobile}, 2009.

\bibitem{hotmobile10sensing}
S.~Saroiu and A.~Wolman.
\newblock I am a sensor, and i approve this message.
\newblock In {\em Proc. of HotMobile}, 2010.

\bibitem{sinai14}
M.~B. Sinai, N.~Partush, S.~Yadid, and E.~Yahav.
\newblock Exploiting social navigation.
\newblock {\em Black Hat Asia}, CoRR:abs/1410.0151, 2015.

\bibitem{SMVHunters:NDSS14}
D.~Sounthiraraj, J.~Sahs, G.~Greenwood, Z.~Lin, and L.~Khan.
\newblock Smv-hunter: Large scale, automated detection of ssl/tls
  man-in-the-middle vulnerabilities in android apps.
\newblock In {\em Proc. of NDSS}, 2014.

\bibitem{Stefanovitch2014}
N.~Stefanovitch, A.~Alshamsi, M.~Cebrian, and I.~Rahwan.
\newblock Error and attack tolerance of collective problem solving: The darpa
  shredder challenge.
\newblock {\em EPJ Data Science}, 3(1):1--27, 2014.

\bibitem{link_talasila10}
M.~Talasila, R.~Curtmola, and C.~Borcea.
\newblock {LINK:} location verification through immediate neighbors knowledge.
\newblock In {\em Proc. of {M}obi{Q}uitous}, 2010.

\bibitem{wowmom09}
F.~Tan, Y.~Borghol, and S.~Ardon.
\newblock Emo: A statistical encounter-based mobility model for simulating
  delay tolerant networks.
\newblock In {\em Proc. of WOWMOM}, 2008.

\bibitem{phone_ccs14}
K.~Thomas, D.~Iatskiv, E.~Bursztein, T.~Pietraszek, C.~Grier, and D.~McCoy.
\newblock Dialing back abuse on phone verified accounts.
\newblock In {\em Proc. of CCS}, 2014.

\bibitem{Tjensvold07}
J.~M. Tjensvold.
\newblock {Comparison of the IEEE 802.11, 802.15.1,802.15.4 and 802.15.6
  wireless standards}.
\newblock
  \url{http://janmagnet.files.wordpress.com/2008/07/comparison-ieee-802-standards.pdf},
  2007.

\bibitem{sumup}
N.~Tran, B.~Min, J.~Li, and L.~Subramanian.
\newblock Sybil-resilient online content voting.
\newblock In {\em Proc. of {NSDI}}, 2009.

\bibitem{sybilroundup-sigcom10}
B.~Viswanath, A.~Post, K.~P. Gummadi, and A.~Mislove.
\newblock An analysis of social network-based sybil defenses.
\newblock In {\em Proc. of {SIGCOMM}}, 2010.

\bibitem{clickstream13}
G.~Wang, T.~Konolige, C.~Wilson, X.~Wang, H.~Zheng, and B.~Y. Zhao.
\newblock You are how you click: Clickstream analysis for sybil detection.
\newblock In {\em Proc. of USENIX Security}, 2013.

\bibitem{sybils-ndss13}
G.~Wang, M.~Mohanlal, C.~Wilson, X.~Wang, M.~Metzger, H.~Zheng, and B.~Y. Zhao.
\newblock Social turing tests: Crowdsourcing sybil detection.
\newblock In {\em Proc. of NDSS}, 2013.

\bibitem{whisperIMC14}
G.~Wang, B.~Wang, T.~Wang, A.~Nika, H.~Zheng, and B.~Y. Zhao.
\newblock Whispers in the dark: Analysis of an anonymous social network.
\newblock In {\em Proc. of IMC}, 2014.

\bibitem{stamp_icnp13}
X.~Wang, J.~Zhu, A.~Pande, A.~Raghuramu, P.~Mohapatra, T.~Abdelzaher, and
  R.~Ganti.
\newblock {STAMP}: Ad hoc spatial-temporal provenance assurance for mobile
  users.
\newblock In {\em Proc. of {ICNP}}, 2013.

\bibitem{sybillimit}
H.~Yu, P.~B. Gibbons, M.~Kaminsky, and F.~Xiao.
\newblock Sybillimit: A near-optimal social network defense against sybil
  attacks.
\newblock In {\em Proc. of IEEE {S{\&}P}}, 2008.

\bibitem{sybilguard}
H.~Yu, M.~Kaminsky, P.~B. Gibbons, and A.~Flaxman.
\newblock Sybilguard: defending against sybil attacks via social networks.
\newblock In {\em Proc. of {SIGCOMM}}, 2006.

\bibitem{foursq}
Z.~Zhang, L.~Zhou, X.~Zhao, G.~Wang, Y.~Su, M.~Metzger, H.~Zheng, and B.~Y.
  Zhao.
\newblock On the validity of geosocial mobility traces.
\newblock In {\em Proc. of HotNets}, 2013.

\bibitem{applaus_zhu2013}
Z.~Zhu and G.~Cao.
\newblock Toward privacy preserving and collusion resistance in a location
  proof updating system.
\newblock {\em IEEE Transactions on Mobile Computing}, 12(1):51--64, 2013.

\end{thebibliography}

\end{document}